\documentclass[twocolumn,showpacs,
aps,superscriptaddress, 
prd,notitlepage,showkeys,
nofootinbib, floatfix]{revtex4-1}

\usepackage[normalem]{ulem}
\usepackage{amssymb}
\usepackage{amsmath}
\usepackage{graphicx}
\usepackage{dcolumn}
\usepackage[colorlinks,urlcolor=blue,citecolor=magenta,linkcolor=blue]{hyperref}
\usepackage{color,units}
\usepackage[dvipsnames]{xcolor} 
\usepackage{lineno}
\usepackage{xspace}
\usepackage{orcidlink}
\usepackage{longtable} 
\usepackage{float} 
\usepackage{multirow}
\usepackage{amsfonts,wasysym,epsfig, verbatim, subfigure, bm,mathrsfs,lipsum}
\usepackage{bm}
\usepackage{soul}
\usepackage{pifont}
\usepackage{tikz}
\usetikzlibrary{shapes.geometric, arrows, shadows, positioning}
\usetikzlibrary{positioning, shapes.misc}

\usepackage{multirow}
\usepackage{longtable,tabu}
\begin{document}

\newcommand{\NITRPHY}{Department of Physics and Astronomy, National Institute of Technology, Rourkela 769008, India}

\newcommand{\BITS}{Department of Physics, Birla Institute of Technology and Science - Pilani, K. K. Birla Goa Campus, NH-17B, Zuarinagar, Sancoale, Goa-403726, India}

\newcommand{\NITRCSE}{Department of Computer Science and Engineering, National Institute of Technology, Rourkela 769008, India}

\title{Decoding the Imprints of Energy-Momentum Squared Gravity in Neutron Stars with Machine Learning Analysis}

\author{Sayantan Ghosh~\orcidlink{0000-0001-8276-1935}}\email{sayantanghosh1999@gmail.com}
\affiliation{\NITRPHY}

\author{Premachand Mahapatra~\orcidlink{0000-0002-3762-8147}}\email{p20210039@goa.bits-pilani.ac.in}
\affiliation{\BITS}

\author{Dipti Deb~\orcidlink{0009-0007-8454-5946}}\email{ddiptideb5@gmail.com}
\affiliation{\NITRCSE}

\date{\today}

\begin{abstract}

Neutron stars (NSs) provide a unique laboratory for testing gravity in the strong-field regime and for searching for deviations from General Relativity (GR). In this work, we investigate the effects of Energy-Momentum Squared Gravity (EMSG) on NS structure and examine whether its signatures can be identified from observable stellar properties using supervised machine learning (ML). We solve the modified Tolman-Oppenheimer-Volkoff equations for approximately $10^{4}$ nuclear equations of state (EOSs) for EMSG coupling parameters $\alpha=\{-5.01,-2.50,0,+2.50,+5.01\}\times10^{-38}\,\mathrm{erg}^{-1}\mathrm{cm}^{3}$, and calculate the gravitational mass $M$, radius $R$, dimensionless tidal deformability $\Lambda$, and fundamental $f$-mode oscillation frequency for each stellar configuration. The free parameter $\alpha$ has a significant effect on $M$, $R$, $\Lambda$ and $f$-mode frequency. Imposing observational constraints on $M$, $R$, and $\Lambda$, we split our datasets into train and test sets, and we employ Random Forest (RF), K-Nearest Neighbors (KNN), Support Vector Machine (SVM), Logistic Regression (LR), and Gaussian Naive Bayes (GNB) to classify the representative sectors $\alpha=\{-5.01,0,+5.01\}\times10^{-38}\,\mathrm{erg}^{-1}\mathrm{cm}^{3}$ using $(M, R,\Lambda,f)$. The RF classifier performs best, achieving an accuracy of approximately $99.85\%$ with precision, recall, and F1-scores exceeding $99.8\%$, while KNN also achieves accuracy above $99\%$. The nearly diagonal confusion matrices demonstrate that the observationally viable NS configurations associated with different EMSG sectors remain highly separable in the multidimensional observable space. Our results show that NS observables retain robust signatures of EMSG even after observational filtering, establishing ML-assisted NS observations as a promising complementary approach for probing modified gravity with current and future multi-messenger observations.\\


\end{abstract}

\maketitle
\section{Introduction}
\label{intro}
Neutron stars (NSs) are extremely compact objects characterized by high density and strong curvature, making them an excellent probe to test gravity in extreme environments \cite{Glendenning, shapiro, Curvature1, Curvature2, Curvature3}. The typical central densities inside the NS exceed nuclear saturation and gravitational potentials, which helps us to probe the strong-field, high-curvature domain far beyond solar-system tests \cite{Kopeikin2014}.

In recent years, multi-messenger observations \cite{Abbott_2017,doi:10.1126/science.aap9811, Abbott_2020, PhysRevX.9.011001, Fragione_2021,sgrb1,sgrb2}, the NICER X-ray telescope's pulse profiles from PSR J0030+0451 \cite{Miller_2019} and the LIGO/Virgo detection of GW170817 – the first binary NS merger – which constrained the binary tidal deformability and common NS radius through the gravitational-wave phasing \cite{GW170817, PhysRevLett.121.091102, PhysRevX.9.011001} provide us with well-structured constraints on NS structure. Combined with precise pulsar mass measurements, these observations yield tight constraints on the NS equation of state (EOS). At the same time, they give us free hand to test gravity itself: NSs and their mergers are sensitive probes of any deviations from General Relativity (GR) in the strong-field regime \cite{Li:2011vx, PhysRevX.11.041050, Miller_2019, Ghosh2025}.

Despite the many successes, GR \cite{Einstein1915} is expected to break down at some scale (e.g., due to dark matter/energy puzzles or singularities) \cite{COPELAND, Riess_1998, Perlmutter_1999, Mahapatra:2024ywx, Liu:2025cwy, Mahapatra:2026utt}, therefore, the alternative theories of gravity are actively explored \cite{CAPOZZIELLO2011167, OLMO20201, NOJIRI201159, NOJIRI20171, CQ, doiNOJIRI, lobo, Sotiriou, OLMO, OLMO20201, Clifton_2012}. Unresolved issues, such as the cosmological constant problem and singularities in gravitational collapse, motivate modifications of the Einstein-Hilbert action \cite{AkarsuNew, EMSG_NAlam, Ghosh2025}. Many extensions introduce new curvature terms (e.g.,\ $f(R)$ gravity \cite{Sotiriou, Astashenok_2015, PhysRevD.93.023501}) or couplings between matter and geometry (e.g.,\ $f(R,T)$ \cite{PhysRevD.84.024020, PhysRevD.97.104041, Mahapatra:2024kfq} or $f(R,T_{\mu\nu}T^{\mu\nu})$ \cite{Katırcı2014, PhysRevD.94.044002} theories). Energy-Momentum Squared Gravity (EMSG), in which the action contains an extra term proportional to the quadratic contraction $T_{\alpha\beta}T^{\alpha\beta}$ of the energy-momentum tensor \cite{PhysRevD.94.044002, AKARSU2023101305, universe10090339}. In this theory, the field equations coincide with GR in vacuum, but acquire extra, density-dependent terms inside matter \cite{PhysRevD.94.044002}. Early studies showed that EMSG can avoid the big-bang singularity (giving a cosmic bounce) and yields a viable cosmological history \cite{PhysRevD.94.044002, PhysRevD.98.024031}. Importantly, since deviations from GR arise only when $T_{\mu\nu}\neq0$, EMSG effects are expected to be most pronounced in the high-curvature cores of compact objects \cite{PhysRevD.98.024031, universe10090339}. Thus, NSs (and black holes (BHs)) are ideal environments to search for EMSG signatures.

There are many works have been carried out on EMSG for compact stars. Nari and Roshan \cite{PhysRevD.98.024031} derived the modified Tolman-Oppenheimer-Volkoff (TOV) equations in EMSG and solved them for a simple polytropic EOS. They found that the NS mass-radius relation can change substantially. Akarsu et al. \cite{EMSG_OAkarsu} solved the hydrostatic equations for realistic EOS and showed that EMSG-induced corrections become ``pronounced in the high-density cores of neutron stars''. Using observed mass–radius measurements, they constrained the EMSG coupling parameter to be extremely small (typically $|\alpha|\lesssim10^{-37}$ in appropriate units) and discussed implications for the hyperon puzzle and early-universe cosmology \cite{EMSG_OAkarsu}. Alam et al. \cite{EMSG_NAlam} performed an extensive survey of NS models in EMSG using a wide range of nuclear EOS. Tidal deformability, compactness, oscillation properties, universal relations, and curvature invariants in EMSG inside the NSs have been explored by Ghosh et al. \cite{Ghosh2025, Ghosh2026JHEAP, Ghosh2026prd}. Since in all studies we are dealing with a huge amount of data and different model parameters, we need some methods which can handle it easily.

Machine learning (ML) methods have become essential in our modern data-driven world,
where finding exact models is often very difficult. ML also has a lot of applications in fundamental physics research. In astrophysics, it has been used to identify signatures of NSs or remnants from various gravitational wave signals \cite{Chatterjee_2020} and supernovae \cite{Abylkairov_2024}, reconstruct gamma-ray
burst (GRB) light curves \cite{dainotti2025}, recognize universal relations among NS properties
\cite{URML}, detect potential dark matter signatures in neutron stars \cite{PrashantML}, analyze the speed of sound in neutron stars \cite{ChatterjeeML} and classify NS EoSs \cite{Gonçalves_2023}.
Due to the complexity of nuclear and particle physics, ML holds significant implications for both theoretical modelling and experimental studies \cite{Hatfield_2021}. The potential of ML tools in nuclear physics has been extensively studied and reviewed in recent work \cite{venneti2026, JiangMLNUPhy}. 

In this study, we have taken 10,000 EOSs and calculated the macroscopic properties of NSs, such as mass, radius, tidal deformability and $f$-mode frequency, by varying the EMSG free parameter $\alpha$. Then, by setting the target values using the observational constraints on mass, radius and tidal deformability, we rearrange the data. After that, we applied different classifiers to classify the data and note down their performance. This is the first study where we have used ML classification techniques to decode the imprints of EMSG in NSs.  

This paper is organized as follows: Sec.~\ref{sec:TF} presents the theoretical formalism. The methodology of applying ML techniques is mentioned in Sec.~\ref{sec:MD}. Sec.~\ref{R&D} contains our results and discussion. We solve the modified TOV equations for each EOS, compute mass-radius relations, etc. and highlight how the EMSG effects depend on the coupling parameter and EOS stiffness. We conclude our discussion in Sec. \ref{con}. Throughout this work, we use a metric signature $(-,+,+,+)$ and adopt geometrized units with $G = c = \hbar = 1$.
\section{Formalism}
\label{sec:TF}
\subsection{Energy-Momentum Squared Gravity (EMSG)}
\label{sec:EMSG}
\textcolor{black}{In Energy-Momentum Squared Gravity (EMSG), the gravitational action is modified by an extra term quadratic in the matter fields. The action can be written as \cite{AkarsuNew}: 
\begin{align}
S=\int \left[\frac{1}{2\kappa}\mathcal{R}+f (\mathcal{L}_{\mathrm{m}}, g_{\mu\nu}T^{\mu\nu}, T_{\mu\nu}T^{\mu\nu})+ \mathcal{L}_{\mathrm{m}}\right]\sqrt{-g}\,\mathrm{d}^4x,
\label{action}
\end{align}
where, $f$ denotes any analytic function of matter-related scalars, such as the matter Lagrangian ($\mathcal{L}_{\mathrm{m}}$) \cite{Harko2010}, the trace ($g_{\mu\nu}T^{\mu\nu}$) \cite{Harko2011}, or the contraction ($T_{\mu\nu}T^{\mu\nu}$) \cite{Katırcı2014,PhysRevD.94.044002,PhysRevD.97.024011,PhysRevD.96.123517}. The Ricci scalar is denoted by $\mathcal{R}$, with gravitational coupling $\kappa =8\pi$. The strength of the EMSG correction is governed by the term $T_{\mu \nu }T^{\mu \nu }$, which is scaled by a real constant parameter $\alpha$.
\\
The energy-momentum tensor ($T_{\mu\nu}$) in terms of $\mathcal{L}_{\mathrm{m}}$, is defined as \cite{EMSG_OAkarsu,EMSG_NAlam}
\begin{align}  \label{tmunudef}
T_{\mu\nu}=-\frac{2}{\sqrt{-g}}\frac{\delta(\sqrt{-g}\mathcal{L}_{\mathrm{m}})}{\delta g^{\mu\nu}}=g_{\mu\nu}\mathcal{L}_{\mathrm{m}}-2\frac{\partial \mathcal{L}_{\mathrm{m}}}{\partial g^{\mu\nu}},
\end{align}
which depends only on the metric tensor components. 
Now we can write the total matter Lagrangian part as \cite{AkarsuNew},
\begin{align}
\mathcal{L}_{\mathrm{m}}^{tot}=\mathcal{L}_{\mathrm{m}}+f.
\end{align}
Putting it in Eq. \eqref{tmunudef}, we get the total $T_{\mu\nu}$ as
\begin{equation}
\begin{aligned}
\label{1tmunudef}
T^{tot}_{\mu\nu}&=-\frac{2}{\sqrt{-g}}\frac{\delta(\sqrt{-g}\mathcal{L}^{tot}_{\mathrm{m}})}{\delta g^{\mu\nu}}\\
&=-\frac{2}{\sqrt{-g}}\frac{\delta(\sqrt{-g}\mathcal{L}_{\mathrm{m}})}{\delta g^{\mu\nu}}-\frac{2}{\sqrt{-g}}\frac{\delta(\sqrt{-g}f)}{\delta g^{\mu\nu}}\,.
\end{aligned}
\end{equation}
Accordingly, we can write 
\begin{align}
\label{Ttot}
    T^{tot}_{\mu\nu} = T_{\mu\nu} + T^{mod}_{\mu\nu},
\end{align}
where 
\begin{align}
\label{tmod}
    T^{mod}_{\mu\nu}=-\frac{2}{\sqrt{-g}}\frac{\delta(\sqrt{-g}f)}{\delta g^{\mu\nu}}.
\end{align}
Similar to Eq. \eqref{tmunudef}, we can express Eq. \eqref{tmod} as \cite{AkarsuNew},
\begin{align}\label{2tmunu}
    T^{mod}_{\mu\nu}=-\frac{2}{\sqrt{-g}}\frac{\delta(\sqrt{-g}f)}{\delta g^{\mu\nu}} = fg_{\mu\nu}-2f_{T^2}\theta_{\mu\nu},
\end{align}
where 
\begin{align}\label{ftheta}
    f_{T^2}=\frac{\partial f}{\partial(T_{\rho\sigma}T^{\rho\sigma})},~~ 
    \theta_{\mu\nu}= \frac{\delta(T_{\rho\sigma}T^{\rho\sigma})}{\delta g^{\mu\nu}}.
\end{align}
In this study, we consider the function $f$ as 
\begin{align}
f(\mathcal{L}_{\mathrm{m}}, g_{\mu\nu}T^{\mu\nu}, T_{\mu\nu}T^{\mu\nu})=\alpha T_{\mu\nu}T^{\mu\nu},
\end{align}
which is the form of EMSG considered in \cite{Ghosh2025,Katırcı2014,PhysRevD.94.044002,PhysRevD.97.024011,PhysRevD.96.123517,EMSG_OAkarsu,EMSG_NAlam}.
Then, Eq.~\eqref{2tmunu} takes the following form
\begin{align}\label{3tmunu}
    T^{mod}_{\mu\nu}=\alpha T_{\rho\sigma}T^{\rho\sigma}g_{\mu\nu}-2\alpha \theta_{\mu\nu}.
\end{align}
Now, by using Eq. \eqref{Ttot}, we can write Einstein's field equations as
\begin{align} \label{Gmunu}
    G_{\mu\nu}=\kappa T_{\mu\nu} + \kappa T^{mod}_{\mu\nu}\,,
\end{align}
where $G_{\mu \nu }=\mathcal{R}_{\mu \nu }-\frac{1}{2}g_{\mu \nu }\mathcal{R}$ is the Einstein tensor. The ideal fluid form of $T_{\mu\nu}$ is given by
\begin{align}  \label{em}
T_{\mu\nu}=(\mathcal{E}+P)u_{\mu}u_{\nu}+P g_{\mu\nu},
\end{align}
where $\mathcal{E} $ and $P$ are the energy density and pressure respectively. $u_{\mu }$ is the four-velocity satisfying the conditions $u_{\mu }u^{\mu }=-1$, $\nabla _{\nu }u^{\mu }u_{\mu }=0$. Substituting into Eq.~\eqref{Gmunu}, the modified Einstein field equations in EMSG take the form
\begin{align}
G_{\mu\nu}=\kappa T_{\mu\nu}+\kappa \alpha \left(g_{\mu\nu}T_{\sigma\epsilon}T^{\sigma\epsilon}-2\theta_{\mu\nu}\right)\,,
\label{fieldeq}
\end{align}
where the new tensor $\theta _{\mu \nu }$ is defined as
\begin{equation}
\begin{aligned} \theta_{\mu\nu}=&~ T^{\sigma\epsilon}\frac{\delta
T_{\sigma\epsilon}}{\delta g^{\mu\nu}}+T_{\sigma\epsilon}\frac{\delta
T^{\sigma\epsilon}}{\delta g^{\mu\nu}} \\ =&-2\mathcal{L}_{\rm
m}\left(T_{\mu\nu}-\frac{1}{2}g_{\mu\nu}T\right)-TT_{\mu\nu} \\
&+2T_{\mu}^{\gamma}T_{\nu\gamma}-4T^{\sigma\epsilon}\frac{\partial^2
\mathcal{L}_{\rm m}}{\partial g^{\mu\nu} \partial g^{\sigma\epsilon}}.
\label{theta} \end{aligned}
\end{equation}
Here, $T = g^{\mu\nu}T_{\mu\nu}$ is the trace of $T_{\mu\nu}$. The final term of \eqref{theta} involves two derivatives of $\mathcal{L}_{\mathrm{m}}$, unlike Eq. \eqref{tmunudef}. For our cases of interest, this term is identically zero (see below); hence, we omit it from further analysis. A similar approach was adopted in \cite{AkarsuNew}.
Taking $\mathcal{L}_{\mathrm{m}}=P$ \cite{Faraoni,PhysRevD.109.104055}, we get the covariant divergence of Eq. \eqref{fieldeq} as
\begin{equation}
\nabla^{\mu}T_{\mu\nu}=-\alpha g_{\mu\nu}\nabla^{\mu}
(T_{\sigma\epsilon}T^{\sigma\epsilon})+2\alpha\nabla^{\mu}\theta_{\mu\nu}.
\label{nonconservedenergy}
\end{equation}
The local conservation of $T_{\mu\nu}$ holds only in the limit $\alpha =0$. Substituting Eq.~\eqref{em} into Eq.~\eqref{theta}, and then inserting the result into Eq. \eqref{fieldeq}, one gets
\begin{eqnarray}
& & G_{\mu\nu}=\kappa \mathcal{E} \left[\left(1+\frac{P}{\mathcal{E}}\right)u_{\mu}u_{\nu}+\frac{P}{\mathcal{E}}g_{\mu\nu}\right] \nonumber \\
&  &+\alpha\kappa\mathcal{E}^2\left[2\left(1+\frac{4P}{\mathcal{E}}+\frac{3P^2}{\mathcal{E}^2}\right)u_{\mu}u_{\nu}+\left(1+\frac{3P^2}{\mathcal{E}^2}\right)g_{\mu\nu}\right].\nonumber\\
\label{fieldeq2}
\end{eqnarray}
Now we can restore the Einstein field equation by redefining the above equation as
\begin{equation}
    G^{\mu\nu} = \kappa T^{\mu\nu}_{\mathrm{eff}},
    \label{EinsteinFEqn}
\end{equation}
where $T^{\mu\nu}_{\mathrm{eff}} = (\mathcal{E}_{\mathrm{eff}}+P_{\mathrm{eff}})u^{\mu}u^{\nu} + P_{\mathrm{eff}}g^{\mu\nu}$, is the effective energy-momentum tensor.
For an ideal fluid, $\mathcal{E}_{\mathrm{eff}}$ is the effective energy density defined as
\begin{equation}
\label{Eeff}
    \mathcal{E}_{\mathrm{eff}} = \mathcal{E} + \alpha\mathcal{E}^2\Bigg(1+\frac{8P}{\mathcal{E}} + \frac{3P^2}{\mathcal{E}^2}\Bigg)\,,
\end{equation}
and $P_{\mathrm{eff}}$ is the effective pressure defined as
\begin{equation}
\label{Peff}
    P_{\mathrm{eff}} = P + \alpha\mathcal{E}^2\Bigg(1+\frac{3P^2}{\mathcal{E}^2}\Bigg)\,.
\end{equation}
As noted in \cite{EMSG_OAkarsu}, the parameter $|\alpha |\sim\mathcal{E} ^{-1}$, and we know that for NSs, $\mathcal{E}\sim 10^{37}\,\mathrm{erg\,cm^{-3}}$ \cite{shapiro}. Consequently, EMSG corrections are expected to become important for compact objects such as NSs when the order of $\alpha$ is approximately $|\alpha |\sim 10^{-37}\,\mathrm{erg^{-1}\,cm^{3}}$.}
\subsection{Modified TOV equations in EMSG}
\label{sec:hydrostatics}
Solving the modified Einstein equations in the EMSG framework leads to the modified TOV equations, given by \cite{Ghosh2025, Ghosh2026prd, Ghosh2026JHEAP, EMSG_NAlam}
\begin{align}
\frac{\mathrm{d} m}{\mathrm{d} r}=4\pi r^2 \mathcal{E} \left[1+\alpha\mathcal{E} \left(1+\frac{8P}{\mathcal{E}}+\frac{3P^2}{\mathcal{E}^2 }\right)\right],  \label{TOV1}
\end{align}
\begin{align}
\frac{\mathrm{d} P}{\mathrm{d} r}&= -\frac{m\mathcal{E} }{r^2}\left(1+\frac{P}{\mathcal{E}}\right) \left( 1-\frac{2m}{r}\right)^{-1}  \notag \\
&\times \left[ 1+\frac{4\pi r^3 P}{m }+\alpha \frac{4\pi r^3\mathcal{E}^2}{m}\left(1+\frac{3P^2}{\mathcal{E}^2}\right)\right] \notag \\
&\times \left[1+2\alpha\mathcal{E}\left(1+\frac{3P}{\mathcal{E}} \right)\right] \left[1 + 2\alpha\mathcal{E} \left(c_s^{-2}+\frac{3P}{\mathcal{E}}\right)\right]^{-1}.  \label{TOV2}
\end{align}
We solved the TOV equations by integrating Eqs.~\eqref{TOV1}-\eqref{TOV2} from $r=0$, where $m(r=0)=0$ and $P(r=0)=P_c$ (the central pressure), upto the stellar surface $r=R$, where $m(r=R)=M$ and $P(r=R)=0$, by specifying a central energy density $\mathcal{E}(r=0)=\mathcal{E}_c$ at the center. This integration was performed for all six EOSs ($P(\mathcal{E})$) by varying the parameter $\alpha$ from negative to positive values to get the mass-radius profile.
\subsection{$f$-mode Oscillations}
\label{fmode}
NSs oscillate when they go through external or internal perturbations, and they emit different mode frequencies \cite{PhysRevD.66.104002}. In this study, we use the Cowling approximation \cite{10.1093/mnras/101.8.367} to solve the fundamental ($f$)-mode nonradial oscillations of spherically symmetric NSs.
\\
The Lagrangian displacement vector of the fluid is given by \\
\begin{equation}
\xi^{i}=\frac{1}{r^2}\Big(e^{-\lambda (r)}W (r),-V
(r)\partial_{\theta},
     -\frac{V(r)}{ \sin^{2}{\theta}}\  \partial _{\phi}\Big)
e^{i\omega t}Y_{lm}(\theta,\phi) ,
\end{equation}
where $Y_{lm}(\theta,\phi)$ represents the spherical harmonics, $\omega$ is the frequency. To determine $\omega$, we have to solve the following system of ordinary differential equations \cite{PhysRevD.83.024014}, which are modified by the effective energy density and pressure in EMSG, represented as \cite{Ghosh2025, GHOSH_FTT}, 
\begin{eqnarray}
\frac{d W(r)}{dr}&=&\frac{d {\cal E}_{\mathrm{eff}}}{dP_{\mathrm{eff}}}\left[\omega^2r^2e^{\lambda
(r)-2\nu (r)}V(r)
+\frac{d \nu(r)}{dr} W (r)\right] \nonumber \\
&&
-l(l+1)e^{\lambda (r)}V (r) \nonumber \\
\frac{d V(r)}{dr} &=& 2\frac{d\nu (r)}{dr} V
(r)-\frac{1}{r^2}e^{\lambda (r)}W (r),
\label{eqn:cowling}
\end{eqnarray}
where $\nu(r)$ and $\lambda(r)$ represent metric functions.
\\
In the close vicinity of the origin, the solution to Eq. (\ref{eqn:cowling}) exhibits the following behaviour:
\begin{equation}
     W (r)=Br^{l+1}, \ V (r)=-\frac{B}{l} r^l,
\label{eq:bc1}
\end{equation}
where $B$ is an arbitrary constant. To ensure that the perturbation pressure becomes zero at the outer boundary of the star's surface, we need to apply the following additional boundary condition,
\begin{equation}
     \omega^2 e^{\lambda (R)-2\nu (R)}V (R)+\frac{1}{R^2}\frac{d\nu
(r)}{dr}\Big|_{r=R}W (R)=0.
\label{eq:bc2}
\end{equation}
Utilizing the boundary conditions outlined in Eq. (\ref{eq:bc1}) and Eq. (\ref{eq:bc2}), we can successfully solve Eq. (\ref{eqn:cowling}) and determine the eigenfrequencies of the PNSs.
\subsection{Tidal Deformability Parameters}
In a binary NS system, one NS experience the external field ($\epsilon_{ij}$) created by its companion star; it acquires a quadrupole moment ($Q_{ij}$). The magnitude of the quadrupole moment is linearly proportional to the tidal field and is given by \citep{Hinderer_2008, Hinderer_2009}
\\
\begin{equation}
Q_{i j}=-\lambda \epsilon_{i j},
\end{equation}
\\
where $\lambda$ is the tidal deformability of a star. $\lambda$ can be defined in terms of tidal Love number $k_{2}$ as $\alpha=\frac{2}{3} k_{2} R^{5}$. The dimensionless tidal deformability of the star is defined as $\Lambda = \lambda / M^{5} = \frac{2}{3} k_{2} C^{-5}$. The GW170817 \citep{GW170817} event constrains $\Lambda_{1.4}$ to be $190_{-120}^{+390}$, while GW190814 \citep{GW190814} put a limit of $\Lambda_{1.4}= 616_{-158}^{+273}$ (in the NS-BH scenario).
\subsection{Equation of State of Nuclear Matter}
In this section we summarize the theoretical framework used to construct the nuclear equation of state (EOS) employed in this work. The EOS is obtained within a relativistic mean-field description of nuclear matter, and the associated model parameters are constrained through a Bayesian inference procedure.
\subsubsection{Relativistic mean-field description}
\label{model}
The properties of dense nuclear matter are described using the relativistic mean-field (RMF) model, where nucleons interact through the exchange of mesonic fields. In this framework the scalar $\sigma$ meson generates an attractive component of the nuclear interaction, while the vector $\omega$ meson produces a short-range repulsive force. The isovector $\varrho$ meson accounts for the neutron–proton asymmetry of the system. 

The effective Lagrangian density for the nucleon field $\Psi$ interacting with these mesons can be written as
\begin{equation}
\begin{aligned}
\mathcal{L} =& 
\bar{\Psi}\Big[\gamma^{\mu}\left(i\partial_{\mu}
-\Gamma_{\omega}A_{\mu}^{(\omega)}
-\Gamma_{\varrho}\boldsymbol{\tau}\cdot\boldsymbol{A}_{\mu}^{(\varrho)}\right) \\
&-(m-\Gamma_{\sigma}\phi)\Big]\Psi
+\frac{1}{2}\left(\partial_{\mu}\phi\partial^{\mu}\phi
-m_{\sigma}^{2}\phi^{2}\right)  \\
&-\frac{1}{4}F_{\mu\nu}^{(\omega)}F^{(\omega)\mu\nu}
+\frac{1}{2}m_{\omega}^{2}A_{\mu}^{(\omega)}A^{(\omega)\mu} \\
&-\frac{1}{4}\boldsymbol{F}_{\mu\nu}^{(\varrho)}
\cdot \boldsymbol{F}^{(\varrho)\mu\nu}
+\frac{1}{2}m_{\varrho}^{2}
\boldsymbol{A}_{\mu}^{(\varrho)}
\cdot \boldsymbol{A}^{(\varrho)\mu},
\end{aligned}
\label{lagrangian}
\end{equation}
where $\gamma^\mu$ denotes the Dirac matrices and $\boldsymbol{\tau}$ are the Pauli matrices in isospin space. The meson field tensors are defined as 
$F^{(\omega,\varrho)\mu\nu}=\partial^\mu A^{(\omega,\varrho)\nu}-\partial^\nu A^{(\omega,\varrho)\mu}$.
The couplings $\Gamma_{\sigma}$, $\Gamma_{\omega}$, and $\Gamma_{\varrho}$ characterize the interaction strengths between nucleons and the corresponding meson fields.

In density-dependent hadronic (DDH) models these couplings vary with the baryon density $\rho$ and are parametrized as
\begin{equation}
\Gamma_M(\rho)=\Gamma_{M,0}\,h_M(x), \qquad x=\rho/\rho_0 ,
\end{equation}
where $\rho_0$ is the nuclear saturation density. For the isoscalar sector we adopt
\begin{equation}
h_M(x)=\exp[-(x^{a_M}-1)],
\label{hm1}
\end{equation}
while the isovector coupling follows
\begin{equation}
h_\varrho(x)=\exp[-a_\varrho(x-1)],
\label{hm2}
\end{equation}
as described in Ref.~\cite{Typel:1999yq}. These density dependencies reproduce the trends obtained in Dirac–Br\"uckner–Hartree–Fock calculations at densities relevant for NS interiors \cite{TerHaar:1986xpv,Brockmann:1990cn,Typel:1999yq}.

Assuming uniform matter and employing the mean-field approximation, the meson fields are replaced by their expectation values. The corresponding field equations become
\begin{eqnarray}
m_\sigma^2\sigma &=& \Gamma_\sigma \bar{\psi}\psi ,\\
m_\omega^2\omega_0 &=& \Gamma_\omega \bar{\psi}\gamma_0\psi ,\\
m_\varrho^2\varrho_3^0 &=& \frac{1}{2}\Gamma_\varrho
\bar{\psi}\gamma_0\tau_3\psi .
\end{eqnarray}

At zero temperature the baryon number density and scalar density are given by
\begin{eqnarray}
\rho &=& \frac{\gamma}{2\pi^2}\sum_{B=p,n}
\int_0^{k_{F_B}} k^2 dk ,\\
\rho_s &=& \frac{\gamma}{2\pi^2}\sum_{B=p,n}
\int_0^{k_{F_B}}
\frac{m^* k^2}{\sqrt{k^2+m^{*2}}} dk ,
\end{eqnarray}
where $k_{F_B}$ denotes the Fermi momentum of nucleon species $B$, $\gamma$ is the spin degeneracy factor, and the effective nucleon mass is $m^*=m-\Gamma_\sigma\sigma$.

The density dependence of the couplings generates an additional rearrangement contribution $\Sigma^r$, which guarantees thermodynamic consistency of the model \cite{Typel:1999yq}. The total energy density of NS matter can then be written as
\begin{eqnarray}
\varepsilon &=&
\frac{1}{\pi^2}
\sum_{B=n,p}\int_0^{k_{F_B}}
k^2\sqrt{k^2+m^{*2}}\,dk \nonumber \\
&+&\frac{1}{2}m_\sigma^2\sigma^2
+\frac{1}{2}m_\omega^2\omega_0^2
+\frac{1}{2}m_\varrho^2(\varrho_3^0)^2
+\varepsilon_{\rm lep},
\end{eqnarray}
where $\varepsilon_{\rm lep}$ represents the leptonic contribution. The pressure follows from the thermodynamic relation
\begin{equation}
P=\sum_{i=n,p,e,\mu}\mu_i\rho_i-\varepsilon .
\end{equation}

The matter inside NS cores is assumed to satisfy $\beta$ equilibrium through the weak interaction processes
\begin{eqnarray}
n &\leftrightarrow& p + e^- + \bar{\nu},\\
n + \nu &\leftrightarrow& p + e^- ,
\end{eqnarray}
and the equilibrium conditions
\begin{equation}
\mu_n=\mu_p+\mu_e, \qquad \mu_e=\mu_\mu .
\end{equation}
Charge neutrality further requires
\begin{equation}
\rho_p=\rho_e+\rho_\mu .
\end{equation}

To obtain a complete NS EOS, the core EOS is matched to the crust description. The outer crust is described by the BPS EOS, while the inner crust is approximated by a polytropic relation $p(\varepsilon)=a_1+a_2\varepsilon^\gamma$ with $\gamma=4/3$ \cite{Carriere:2002bx}. The uncertainties associated with this matching procedure have been discussed in Refs.~\cite{Fortin:2016hny,Pais:2016xiu,Lopes:2020xlf, Rather:2020gja}.

The EOS of asymmetric nuclear matter can be expanded around the symmetric limit as
\begin{equation}
\epsilon(\rho,\delta)
\simeq
\epsilon(\rho,0)+S(\rho)\delta^2 ,
\label{eq:eden}
\end{equation}
where $\delta=(\rho_n-\rho_p)/\rho$ denotes the isospin asymmetry. Expanding around the saturation density $\rho_0$ introduces bulk nuclear matter parameters such as the incompressibility $K_0$, skewness $Q_0$, and higher-order derivatives that characterize the density dependence of symmetric matter and the symmetry energy.

\subsubsection{Bayesian inference of model parameters}
\label{bayes}

The parameters of the RMF model are constrained using Bayesian statistical inference, which provides a systematic framework for updating parameter probabilities in light of available data \cite{Wesolowski:2015fqa, Furnstahl:2015rha, Ashton:2018jfp, Landry:2020vaw}. Bayes' theorem relates the posterior probability distribution of the parameter vector $\boldsymbol{\theta}$ to the prior information and the likelihood function,
\begin{equation}
P(\boldsymbol{\theta}|D)=
\frac{\mathcal{L}(D|\boldsymbol{\theta})P(\boldsymbol{\theta})}
{\mathcal Z},
\label{eq:bt}
\end{equation}
where $D$ represents the dataset, $P(\boldsymbol{\theta})$ denotes the prior distribution, and $\mathcal Z$ is the Bayesian evidence.

Assuming Gaussian errors, the likelihood function takes the form
\begin{equation}
\mathcal{L}(D|\boldsymbol{\theta})
=
\prod_j
\frac{1}{\sqrt{2\pi\sigma_j^2}}
\exp\!\left[
-\frac{1}{2}
\left(
\frac{d_j-m_j(\boldsymbol{\theta})}{\sigma_j}
\right)^2
\right],
\label{eq:likelihood}
\end{equation}
where $d_j$ and $m_j(\boldsymbol{\theta})$ denote the observed data and corresponding model predictions, respectively.

Posterior distributions for individual parameters are obtained by marginalizing over the remaining parameters,
\begin{equation}
P(\theta_i|D)
=
\int P(\boldsymbol{\theta}|D)
\prod_{k\neq i} d\theta_k .
\label{eq:mpd}
\end{equation}

To sample the posterior distribution we employ nested sampling algorithms \cite{Ashton:2022grj}. In particular, we use the \texttt{Bilby} inference framework \cite{Ashton:2018jfp} with the \texttt{PyMultiNest} \cite{Buchner_2023} and \texttt{Dynesty} \cite{Speagle_2020} samplers to efficiently explore the multidimensional parameter space and evaluate the Bayesian evidence.

\section{Methodology}
\label{sec:MD}

\subsection{Overview of the EMSG--Machine Learning Framework}

In this work, we develop a physics-informed machine learning (ML) framework to investigate whether macroscopic NS observables can encode signatures of modified gravity within the Energy--Momentum Squared Gravity (EMSG) theory. In EMSG, the coupling parameter $\alpha$ introduces additional matter-dependent corrections to the stellar structure equations, thereby modifying the internal pressure support and global NS properties.

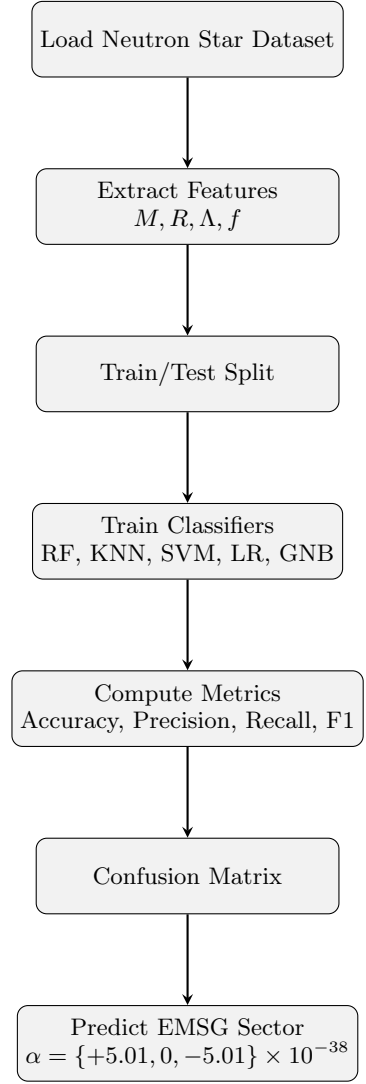
\begin{figure}[htbp]
\centering

\begin{tikzpicture}[
node distance=12mm,
every node/.style={font=\small},
box/.style={
    rectangle,
    rounded corners,
    minimum width=40mm,
    minimum height=10mm,
    draw=black,
    fill=gray!10
},
arrow/.style={thick,->,>=stealth}
]

\node[box] (load) {Load Neutron Star Dataset};
\node[box, below=of load, align=center] (features) {Extract Features\\$M, R, \Lambda, f$};
\node[box, below=of features] (split) {Train/Test Split};
\node[box, below=of split, align=center] (train) {Train Classifiers\\RF, KNN, SVM, LR, GNB};
\node[box, below=of train,
align=center] (metrics) {Compute Metrics\\Accuracy, Precision, Recall, F1};
\node[box, below=of metrics,
align=center](cm) {Confusion Matrix};
\node[box, below=of cm,
align=center] (predict) {Predict EMSG Sector\\$\alpha = \{+5.01,0,-5.01\}\times10^{-38}$};

\draw[arrow] (load) -- (features);
\draw[arrow] (features) -- (split);
\draw[arrow] (split) -- (train);
\draw[arrow] (train) -- (metrics);
\draw[arrow] (metrics) -- (cm);
\draw[arrow] (cm) -- (predict);

\end{tikzpicture}

\caption{Flowchart summarizing the machine learning pipeline used to classify 
neutron stars in Energy--Momentum Squared Gravity (EMSG).}
\label{fig:ml_flowchart}
\end{figure}

The complete workflow, consistent with Fig.~\ref{fig:ml_flowchart} and the Results section, consists of two stages:
\begin{enumerate}
    \item Identification of observationally viable NS configurations
    \item Classification of the underlying EMSG coupling parameter $\alpha$
\end{enumerate}

This two-step strategy ensures that the ML analysis remains grounded in physically realistic NS models.

\subsection{Dataset Generation and Feature Space}

The dataset is constructed by solving the modified Tolman--Oppenheimer--Volkoff (TOV) equations in EMSG for approximately $10^4$ nuclear equations of state (EOSs). For each EOS, stellar configurations are generated over a wide range of central energy densities and for different values of the coupling parameter:
\begin{equation}
\alpha=\{-5.01,-2.50,0,+2.50,+5.01\}\times10^{-38}\,\mathrm{erg^{-1}\,cm^{3}}.
\end{equation}

For every equilibrium configuration, we compute the following macroscopic observables:
\begin{equation}
\mathbf{x} = (M,\, R,\, \Lambda,\, f),
\end{equation}
where each quantity has direct physical significance:
\begin{itemize}
    \item $M, R$: global structural properties governed by hydrostatic equilibrium
    \item $\Lambda$: tidal response sensitive to stellar compactness
    \item $f$: oscillation frequency determined by average density
\end{itemize}

These features collectively capture the modifications induced by the EMSG parameter $\alpha$ on NS structure.

\subsection{Feature Selection and Physical Labeling} \subsubsection{Stage I: Observational Viability Filtering}

The first stage of the ML pipeline consists of identifying NS configurations that are consistent with current astrophysical observations. Instead of using only the mass–radius relation, we impose a set of physically motivated constraints involving multiple observables~\cite{Antoniadis:2013pzd, Fonseca:2021wxt, Romani:2022jhd, Miller:2019cac, Riley:2019yda, Miller:2021qha, Riley:2021pdl, LIGOScientific:2017vwq, LIGOScientific:2018cki}:

\begin{equation}
M > 2\,M_\odot, \quad R > 12~\mathrm{km}, \quad \Lambda > 580.
\end{equation}

These conditions are motivated by:
\begin{itemize}
    \item Massive pulsar measurements ($M \gtrsim 2\,M_\odot$) ~\cite{Antoniadis:2013pzd, Fonseca:2021wxt, Romani:2022jhd}.
    \item NICER radius constraints ($R \sim 12$--13 km)~\cite{Miller:2019cac, Riley:2019yda, Miller:2021qha, Riley:2021pdl}.
    \item Gravitational-wave bounds on tidal deformability (GW170817)~\cite{LIGOScientific:2017vwq, LIGOScientific:2018cki}.
\end{itemize}

Based on these criteria, each configuration is assigned a binary label:
\begin{equation}
y =
\begin{cases}
1, & \text{observationally viable neutron star}, \\
0, & \text{excluded by observations}.
\end{cases}
\end{equation}

This step is crucial, as it removes physically unrealistic models and restricts the analysis to the observationally allowed region of the parameter space.

\subsubsection{Stage II: Classification of EMSG Coupling Parameter}

After filtering, only the configurations satisfying $y=1$ are retained. The second stage of the ML framework aims to determine whether the remaining macroscopic observables retain sufficient information to distinguish the underlying EMSG coupling parameter $\alpha$.

We therefore perform a supervised multi-class classification using three representative sectors:
\begin{equation}
\alpha=
\left\{
+5.01\times10^{-38},\;
0,\;
-5.01\times10^{-38}
\right\}
\ \mathrm{erg^{-1}\,cm^{3}}.
\end{equation}

The ML task is thus defined as:
\begin{equation}
\mathbf{x} \rightarrow \alpha,
\end{equation}
where the input features correspond to observationally viable NS properties.

This formulation directly tests whether EMSG-induced modifications to stellar structure produce separable signatures in the multidimensional observable space.

\subsubsection{Stage III: Data Preprocessing: Scaling and Splitting}

Before training the machine learning classifiers, the generated NS dataset is subjected to a sequence of preprocessing steps to ensure that the input data are suitable for classification. The main steps are: 

\begin{itemize}
    \item \textbf{Data cleaning:} 
The dataset is examined for missing, non-finite, or otherwise unsuitable entries in the NS observables. Only physically meaningful configurations are retained for the subsequent analysis. 

\item \textbf{Feature scaling:} The input observables $(M,R,\Lambda,f)$ have different numerical ranges and units. Therefore, the features are scaled before applying the machine learning algorithms so that variables with larger numerical values do not artificially dominate the classification. 

\item \textbf{Class balancing:} For the first-stage viability classification, the imbalance between observationally viable and non-viable NS configurations is mitigated to prevent the classifier from being biased toward the majority class. 

\item \textbf{Train--test split:} The processed dataset is divided into training and testing subsets. A stratified split is used so that the relative representation of the EMSG classes is preserved in both subsets.

\end{itemize}

\subsection{Classification Models}
We employ five supervised machine learning algorithms to classify NS configurations:

\begin{itemize}
    \item \textbf{Random Forest (RF):} Captures non-linear correlations among observables and is highly effective in detecting EMSG-induced structural variations \cite{verma2016detection}.
    
    \item \textbf{k-Nearest Neighbours (KNN):} Classifies configurations based on proximity in feature space, naturally identifying clustering patterns in $(M, R, \Lambda)$ \cite{Cover1967NearestNP}.
    
    \item \textbf{Support Vector Machine (SVM):} Constructs optimal separating boundaries in high-dimensional space, useful for subtle differences induced by $\alpha$ \cite{Cortes1995SupportVectorN}.
    
    \item \textbf{Logistic Regression (LR):} Provides a baseline linear classifier, useful for assessing the degree of separability in the dataset \cite{Foster2018LogisticRT}.
    
    \item \textbf{Gaussian Naive Bayes (GNB):} A probabilistic model assuming Gaussian-distributed features, serving as a simple benchmark \cite{Raschka2014NaiveBA, Islam2021GGNBGG}.
\end{itemize}

\subsubsection{Model Training and Evaluation}

Each classifier is trained using the training subset and subsequently evaluated on the independent test subset, which contains configurations not used during model training. The classification performance is assessed using four standard metrics: accuracy, precision, recall, and F1-score. These metrics provide complementary measures of how reliably the models distinguish NS configurations belonging to the three EMSG coupling sectors.

\begin{figure*}[htbp]
    \centering
    \includegraphics[width=0.497\linewidth]{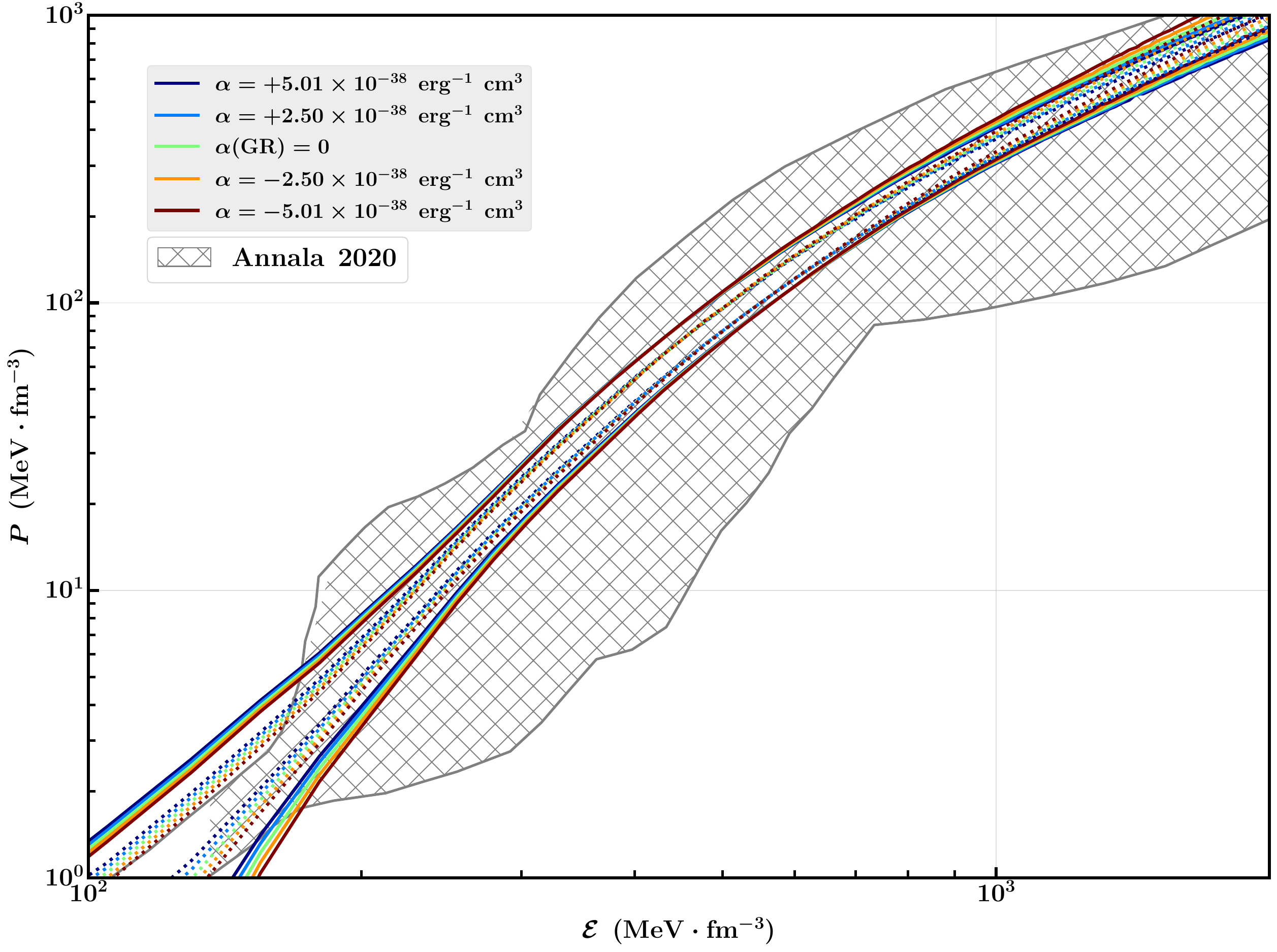}
    \includegraphics[width=0.497\linewidth]{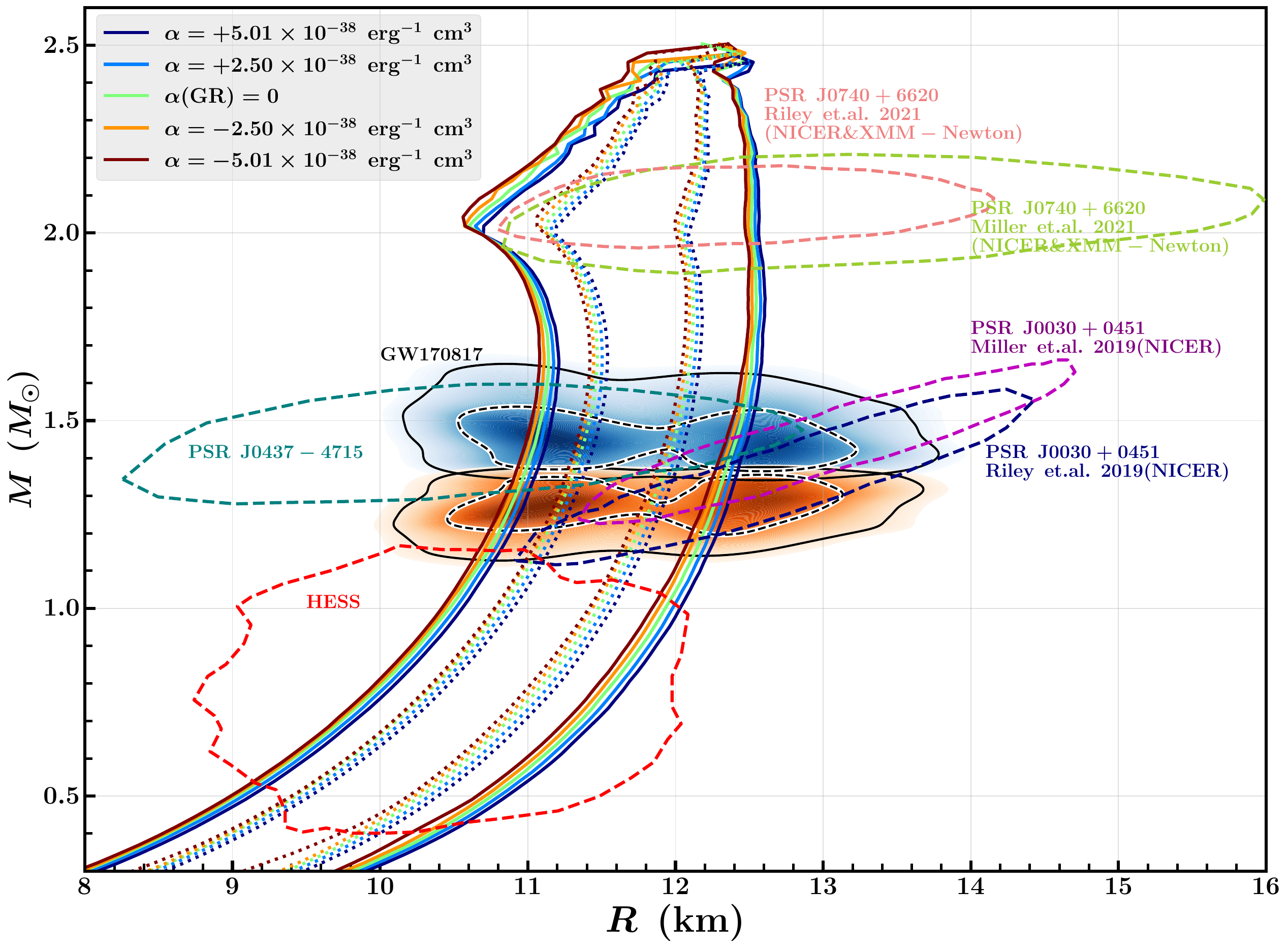}
    \caption{\textit{The left panel} shows the pressure ($P$) as a function of energy density ($\mathcal{E}$) for the representative EOS together with the modifications introduced by different values of the EMSG coupling parameter, $\alpha$. The shaded region (or black band) represents the observationally constrained EOS from Annala (2020).
\textit{The right panel} presents the corresponding NS mass-radius ($M$--$R$) relation. The figure demonstrates that EMSG can produce NS configurations consistent with current astrophysical observations while exhibiting measurable deviations from the predictions of General Relativity. The solid and dotted lines are for the 90\% and 50\% intervals, respectively.}
    \label{EOS&MR}
\end{figure*}

For a given EMSG sector, the \textit{accuracy} measures the fraction of all test configurations that are assigned to the correct class,
\begin{equation}
\mathrm{Accuracy}
=
\frac{TP+TN}{TP+TN+FP+FN}
\end{equation}
where $TP$ and $TN$ denote the numbers of correctly classified positive and negative samples, known as True Positive and True Negative, while $FP$ and $FN$ denote the corresponding misclassified samples, known as False Positive and False Negatively respectively. Since our problem involves three EMSG sectors, the accuracy is calculated over all classes in the multi-class test set.

The \textit{precision} measures the reliability of the predicted class labels and is defined as
\begin{equation}
\mathrm{Precision}
=
\frac{TP}{TP+FP}
\end{equation}
where a high precision indicates that configurations predicted to belong to a particular EMSG sector are rarely assigned to that sector incorrectly. The \textit{recall} measures the ability of the classifier to correctly identify configurations that truly belong to a given EMSG sector,
\begin{equation}
\mathrm{Recall}
=
\frac{TP}{TP+FN}
\end{equation}
Thus, a high recall indicates that only a small fraction of configurations belonging to that EMSG sector are missed by the classifier. The \textit{F1-score} combines precision and recall through their harmonic mean,
\begin{equation}
\mathrm{F1 \, score}
=
2 \times \left( \frac{\mathrm{Precision}\times\mathrm{Recall}}
{\mathrm{Precision}+\mathrm{Recall}} \right)
\end{equation}
and therefore provides a single measure of the balance between correct class assignment and the identification of all configurations belonging to each EMSG sector.

For the multi-class classification considered here, these quantities are evaluated for each EMSG sector and summarized together with the overall accuracy. In addition, we use the confusion matrix to examine the class-wise predictions in greater detail. Each element of the confusion matrix represents the number of test configurations belonging to a true EMSG sector that are assigned to a predicted sector. Correct classifications appear along the diagonal, whereas off-diagonal elements represent misclassifications between different EMSG coupling sectors. Consequently, a nearly diagonal confusion matrix indicates that the macroscopic observables $(M,R,\Lambda,f)$ provide a well-separated feature space for distinguishing the considered EMSG sectors.

The combination of these evaluation measures allows us to assess not only the overall predictive performance of each classifier, but also whether the observed separability is maintained for each individual EMSG coupling sector. The resulting accuracy, precision, recall, F1-score, and confusion matrices are presented and discussed in Sec.~\ref{R&D}.

\begin{figure*}[htbp]
    \centering
    \includegraphics[width=0.497\linewidth]{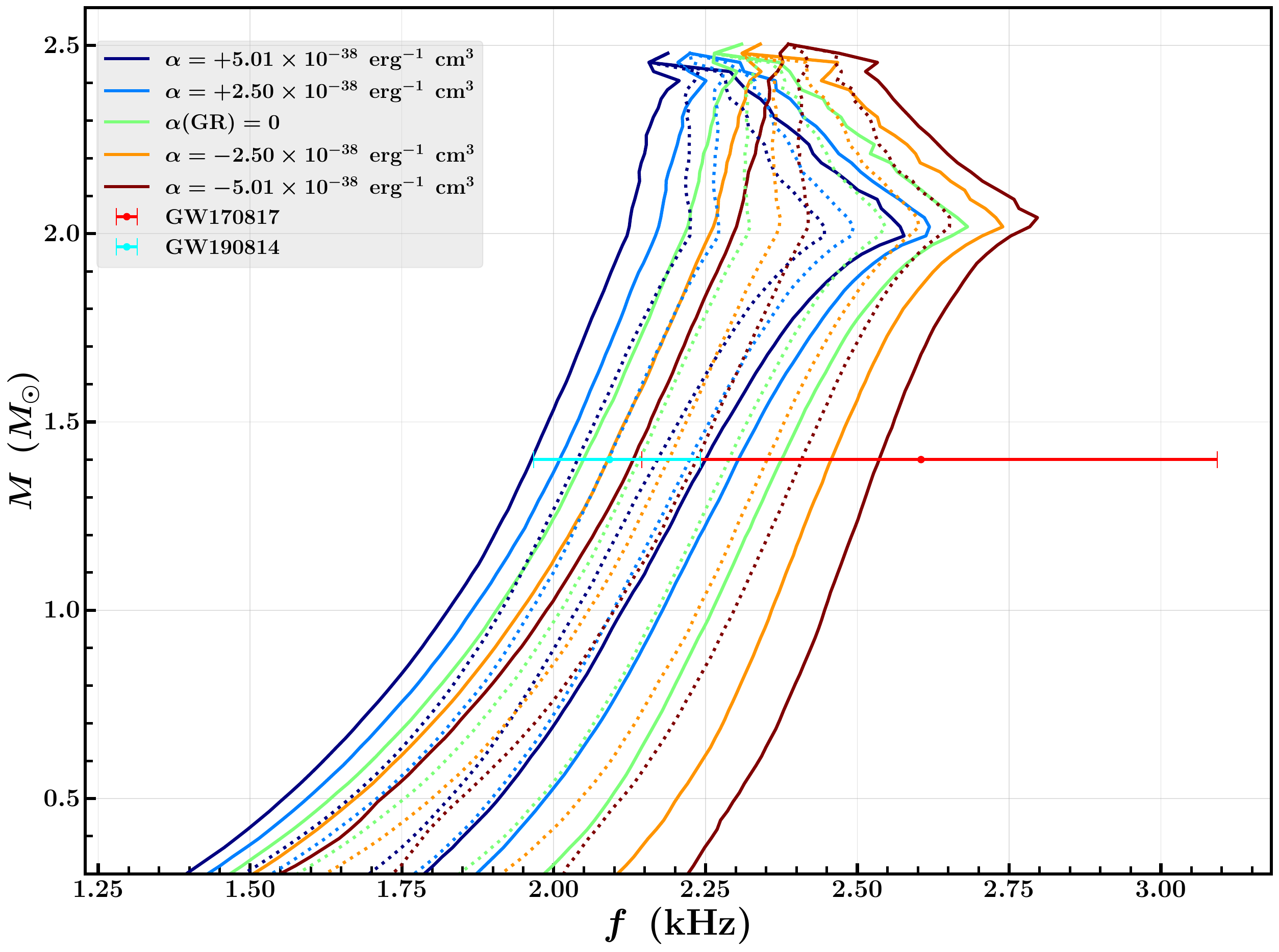}
    \includegraphics[width=0.497\linewidth]{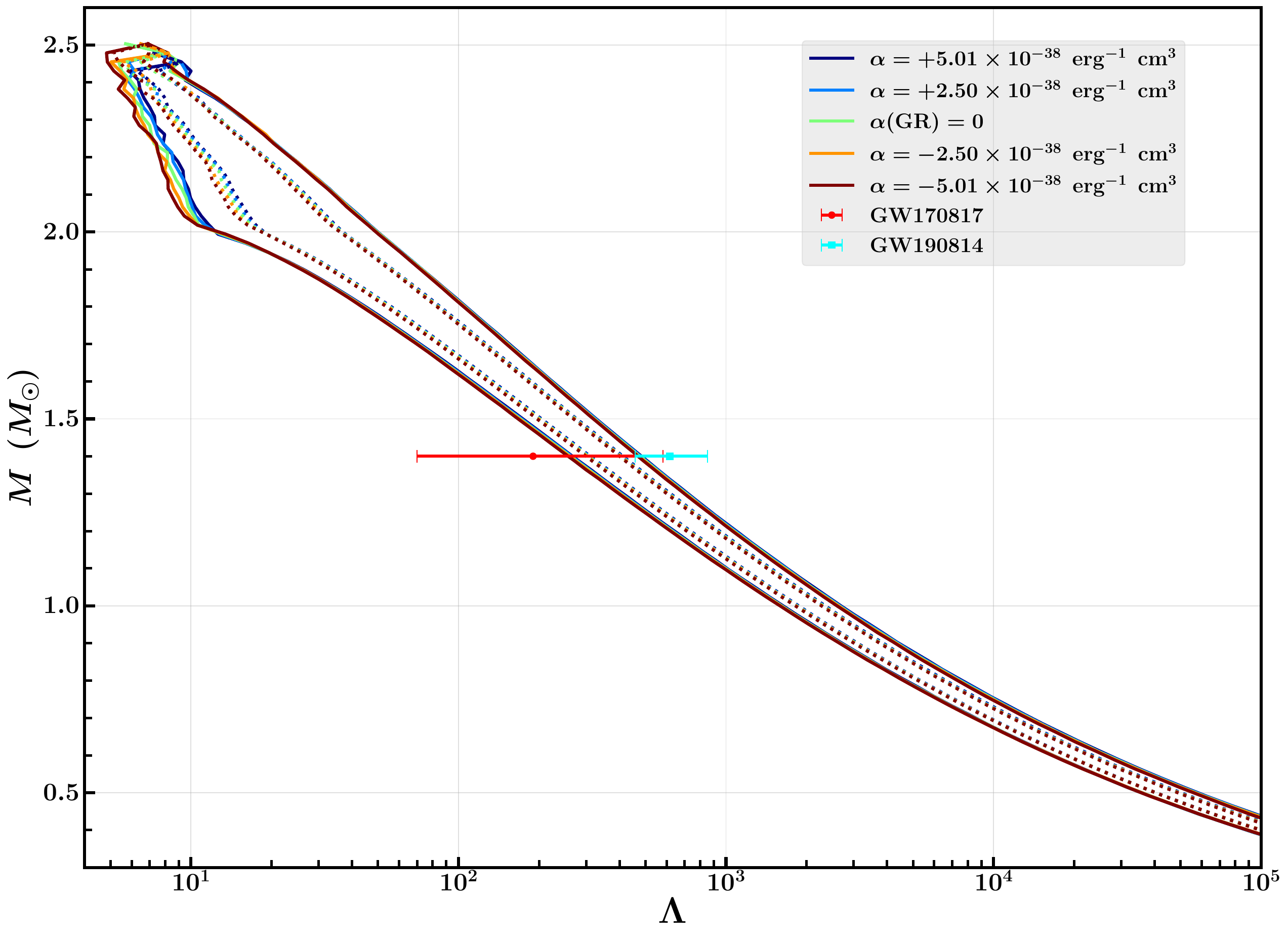}
    \caption{\textit{The left panel} represents the $f$-mode frequency variation with mass, and \textit{the right panel} shows the tidal deformability variation with mass. The solid and dotted lines are for the 90\% and 50\% intervals, respectively.}
    \label{fLambda}
\end{figure*}
\section{Results and Discussion}
\label{R&D}

In this section, we present our comprehensive analysis of this work.
\subsection{Neutron Stars properties in Energy-Momentum Squared Gravity}

To investigate the effects of Energy-Momentum Squared Gravity (EMSG) on NS observables, we generated a comprehensive dataset by solving the modified Tolman-Oppenheimer-Volkoff (TOV) equations for approximately $10,000$ nuclear equations of state (EOSs). For each EOS, equilibrium stellar configurations were computed over a wide range of central energy densities for different values of the EMSG coupling parameter,
\\
$\alpha=\{-5.01,-2.50,0,+2.50,+5.01\}\times10^{-38}\,
\mathrm{erg^{-1}\,cm^{3}}$.
\\
The corresponding macroscopic observables, namely the gravitational mass ($M$), radius ($R$), tidal deformability ($\Lambda$), and the fundamental ($f$)-mode oscillation frequency, were calculated for every stellar configuration.
\\
Fig.~\ref{EOS&MR} presents the pressure-energy density relation together with the corresponding mass-radius relation for different values of the EMSG coupling parameter. The quadratic matter corrections introduced in EMSG modify the effective energy density and pressure, thereby changing the stellar hydrostatic equilibrium. Positive values of $\alpha$ provide additional pressure support inside the stellar core, enabling NSs to sustain larger maximum masses with relatively larger radii compared to General Relativity (GR). Conversely, negative values of $\alpha$ reduce the effective pressure support, producing more compact stellar configurations with smaller maximum masses.
\\
The predicted mass--radius relations are compared with observational constraints obtained from NICER measurements of PSR J0030+0451 and PSR J0740+6620 together with the gravitational-wave constraints from GW170817. Several EMSG models, particularly those corresponding to positive values of $\alpha$, satisfy these observational limits while exhibiting measurable deviations from GR. These results indicate that EMSG remains compatible with present astrophysical observations over an appropriate range of the coupling parameter.

\begin{table*}[htbp]
\caption{Accuracy, Precision, Recall and F1-Score values are presented for different classifier for different values of $\alpha$.}
\centering
\setlength{\tabcolsep}{3.4pt}
\renewcommand{\arraystretch}{1.8}
\scalebox{0.88}{
    \begin{tabular}{cccccccccccccccc}
        \hline \hline
        & \multicolumn{3}{c}{Accuracy} && \multicolumn{3}{c}{Precision}&& \multicolumn{3}{c}{Recall} && \multicolumn{3}{c}{F1-Score} \\
            \cline {2-4} \cline {6-8} \cline {10-12} \cline {14-16}
        $\alpha (10^{-38}) =$ & $+5.01$  & 0  & $-5.01$ && $+5.01$ &  0  &  $-5.01$ && $+5.01$ &  0  &  $-5.01$ && $+5.01$ &  0  &  $-5.01$ \\
        \hline
        RF &0.998525  &0.998754   &0.999081   &&0.999686  &0.999911  &0.999953   &&0.997363   &0.997596    &0.998209    &&0.998523    &0.998752    &0.999080    \\
        KNN &0.992156  &0.994680   &0.994555   &&0.988900  &0.992291  &0.992627   &&0.995486   &0.997106    &0.996511    &&0.992182    &0.994693    &0.994566    \\
        SVM &0.968669  &0.969327   &0.972823   &&0.956072  &0.954712  &0.956197   &&0.982480   &0.985398    &0.991043    &&0.969096    &0.969812    &0.973308    \\
        LR &0.967418  &0.966011   &0.970442   &&0.956124 &0.952338  &0.953138   &&0.979798   &0.981125    &0.989534    &&0.967816    &0.966517    &0.970995    \\
        GNB &0.918276  &0.926969   &0.935109   &&0.860984  &0.874590  &0.887257   &&0.997631   &0.996884    &0.996889    &&0.924284    &0.931741    &0.938883    \\
        \hline \hline
    \end{tabular}}
    \label{tab:APRF}
\end{table*}

\subsection{Tidal deformability, oscillation frequency}

The modifications introduced by EMSG also affect several secondary NS observables. Fig.~\ref{fLambda} shows the variation of the fundamental ($f$)-mode oscillation frequency and the dimensionless tidal deformability with stellar mass for different values of the coupling parameter.
\\
Since the $f$-mode frequency primarily depends on the average density and compactness of the star, changes in the internal matter distribution produced by EMSG naturally modify the oscillation spectrum. Positive values of $\alpha$ generally shift the oscillation frequencies relative to the GR prediction through their influence on stellar compactness, while negative values produce the opposite behaviour. Such deviations may become observable through future gravitational-wave observations of oscillating neutron stars.
\\
The tidal deformability exhibits a similarly systematic dependence on the coupling parameter. Because $\Lambda \propto C^{-5}$, even modest changes in compactness produce significant differences in tidal deformability. Positive values of $\alpha$ generally increase $\Lambda$, whereas negative values decrease it. The resulting sequences remain consistent with the constraints obtained from GW170817 and GW190814 over the considered parameter range, suggesting that tidal deformability provides an efficient probe of EMSG corrections.

\begin{figure*}
    \centering
    \includegraphics[width=0.3\linewidth]{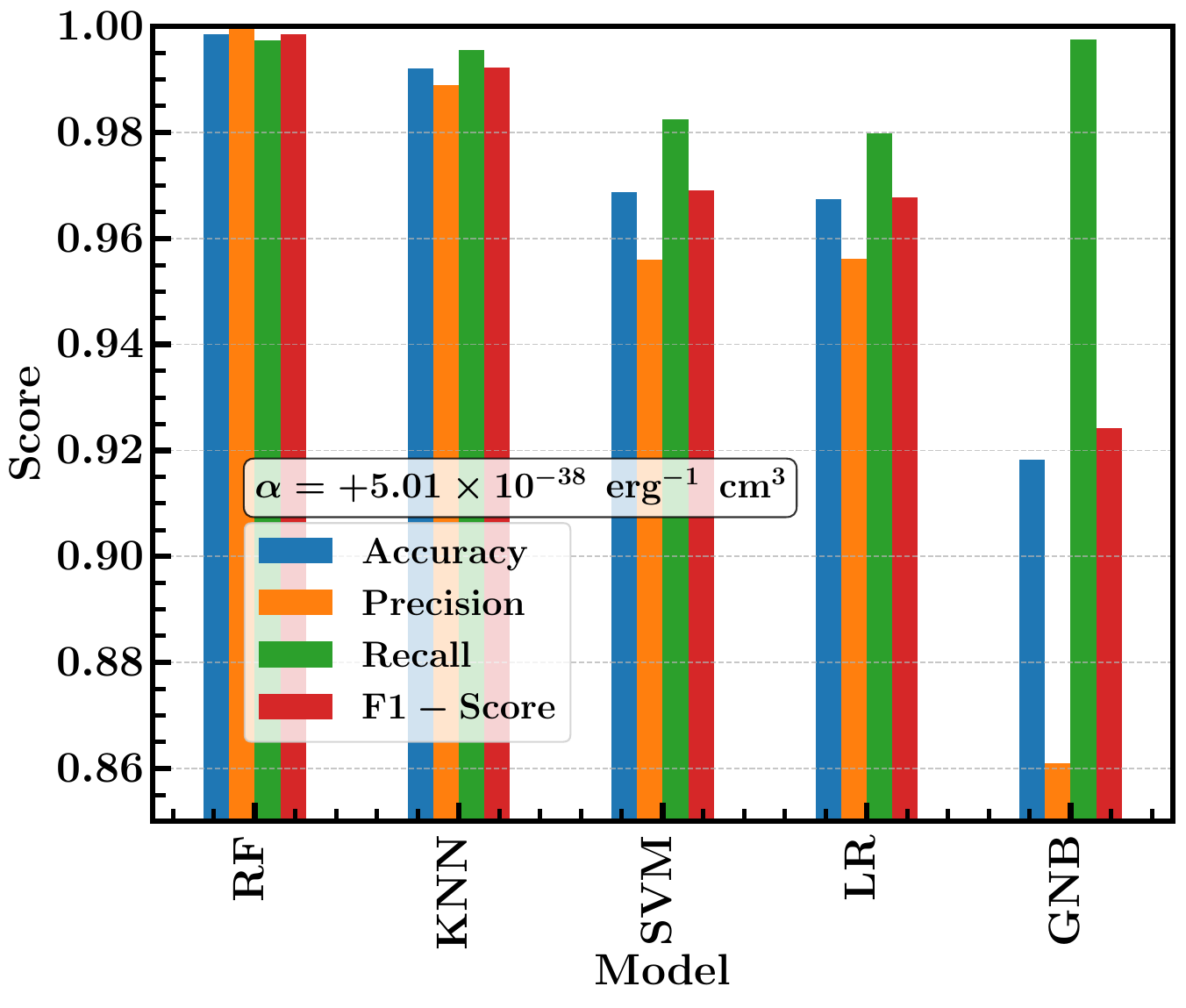}
    \includegraphics[width=0.3\linewidth]{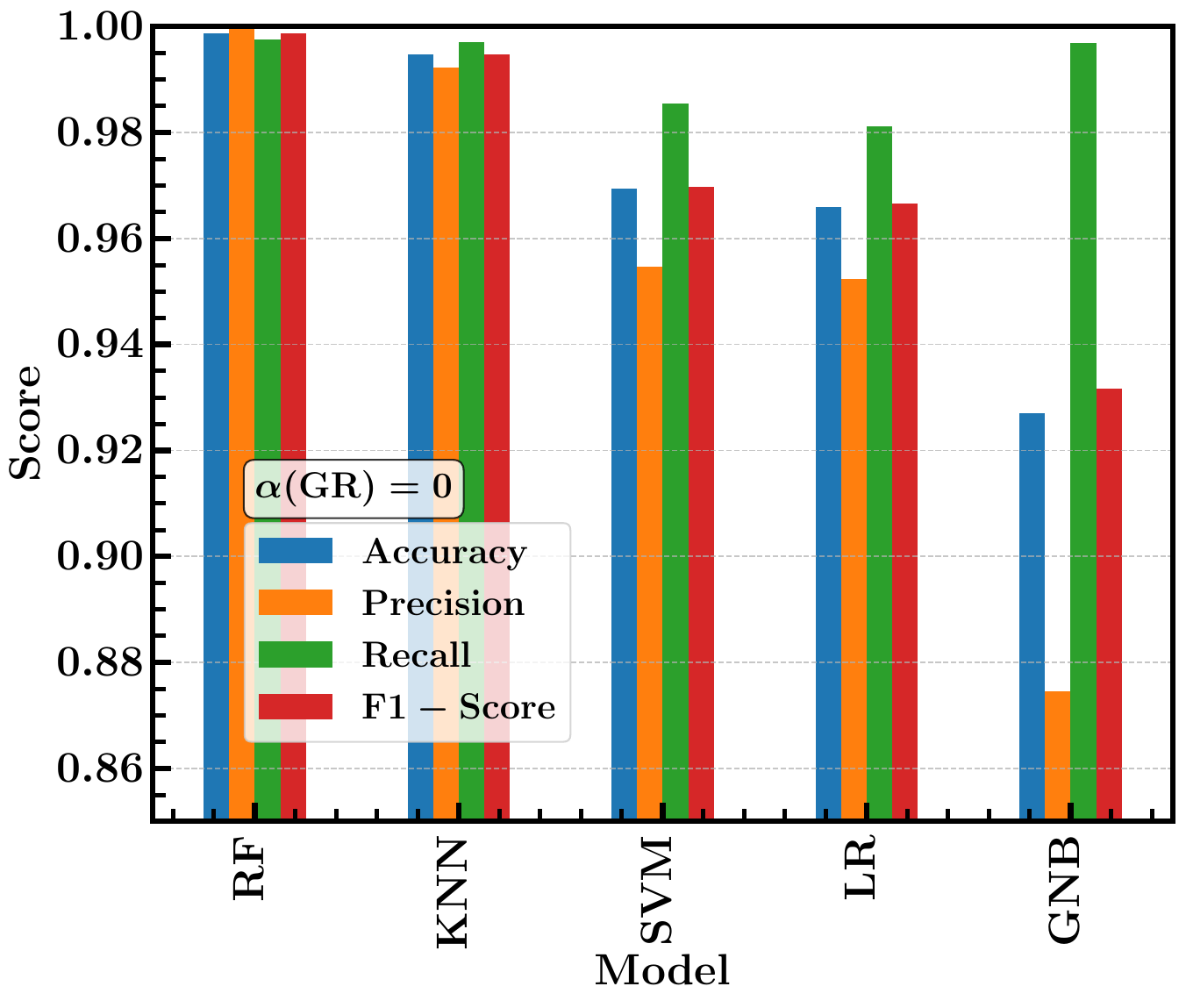}
    \includegraphics[width=0.3\linewidth]{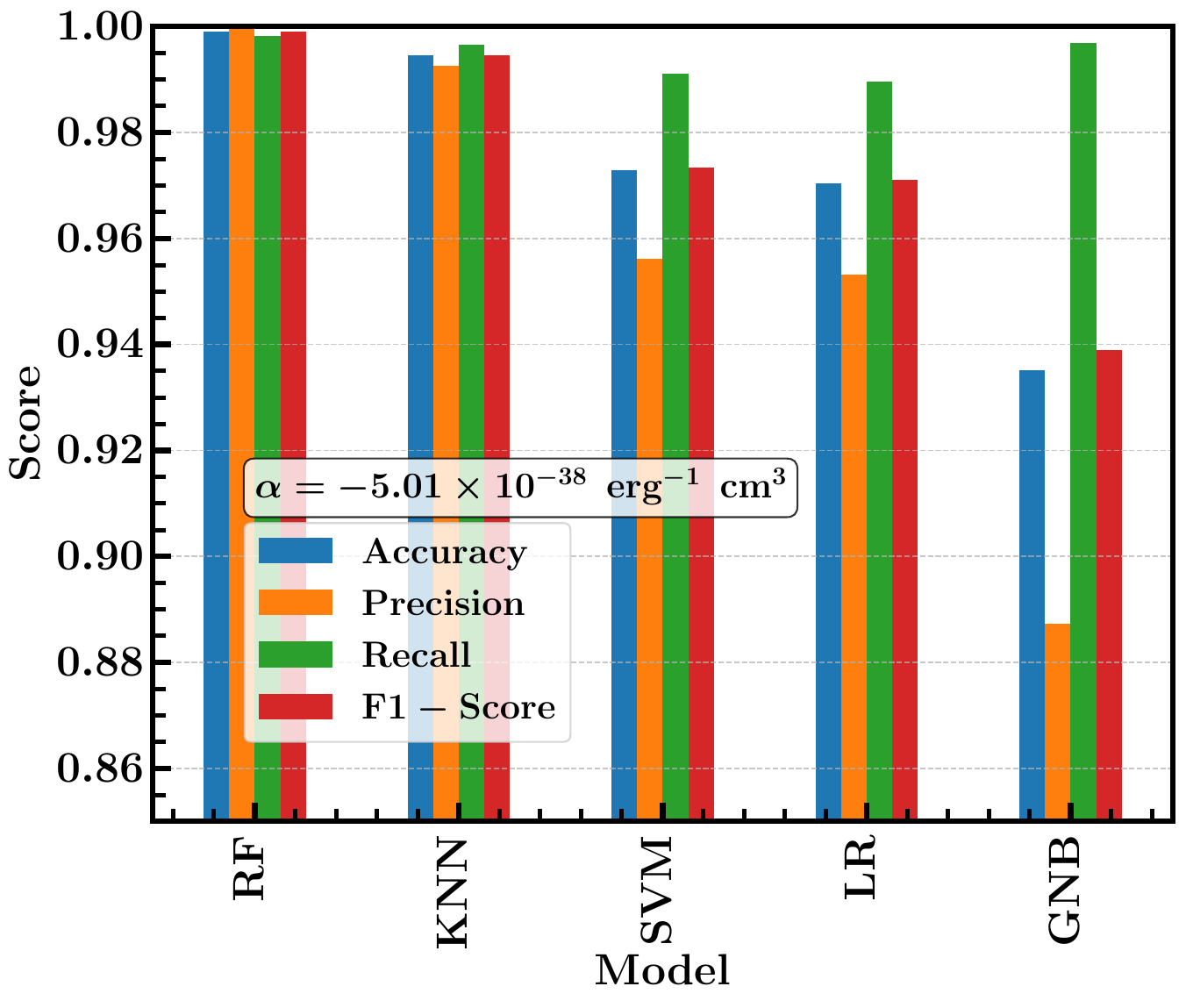}
    \caption {Scores of different classifiers are represented, where different colour bars represent Accuracy, Precision, Recall and F1-Score values.}
    \label{fig:APRF}
\end{figure*}
\begin{figure*}
    \centering
    \includegraphics[width=0.3\linewidth]{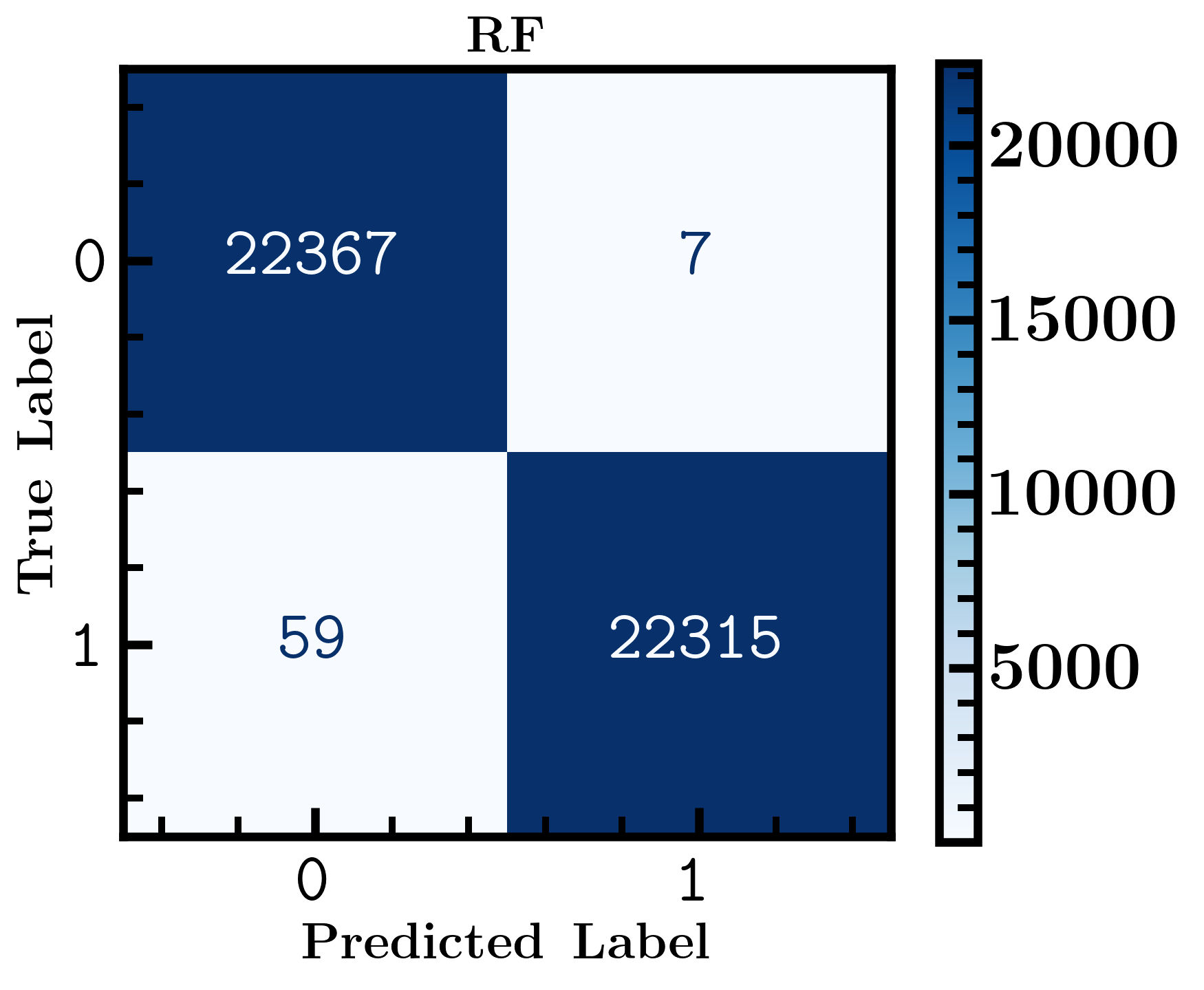}
    \includegraphics[width=0.3\linewidth]{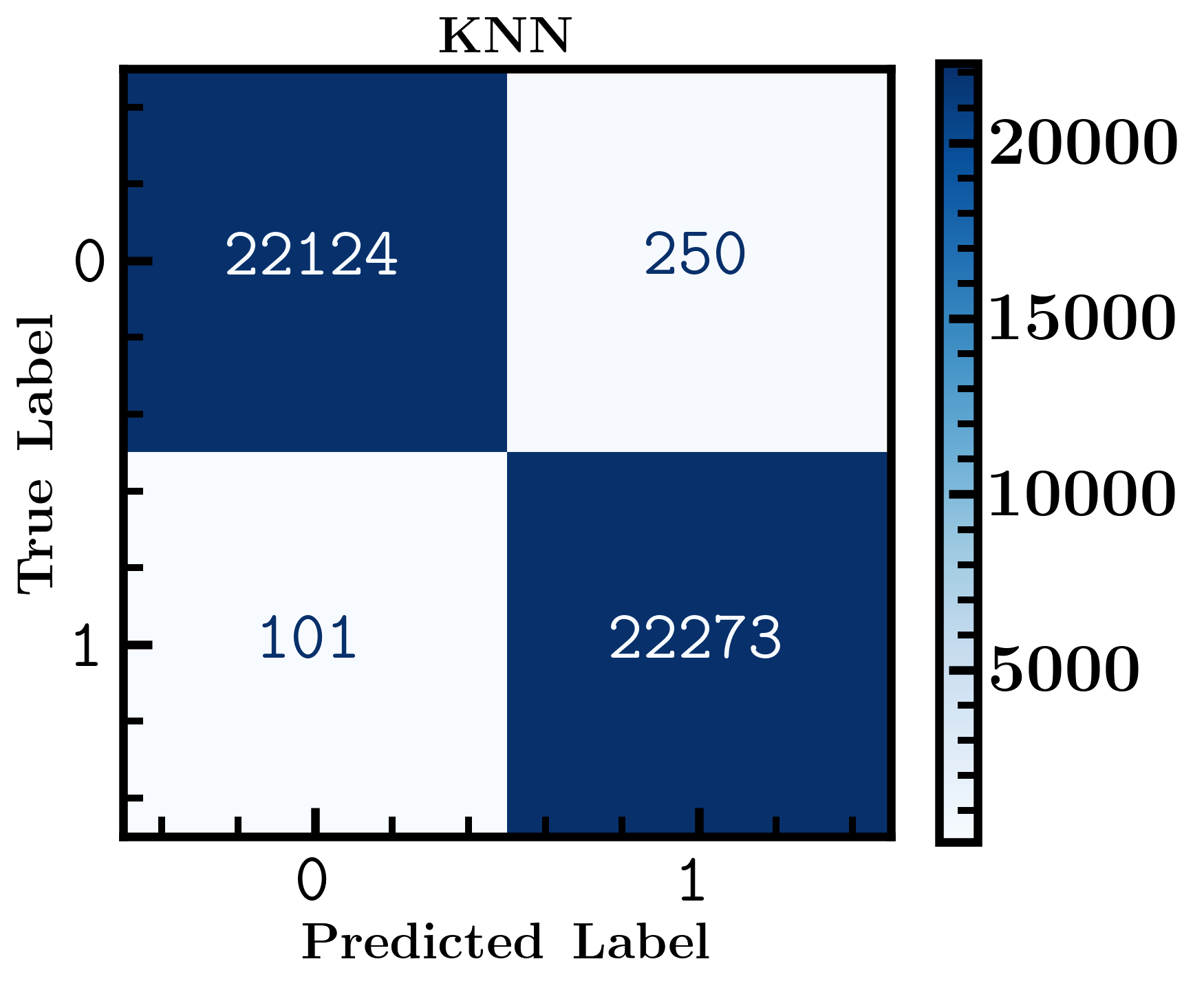}
    \includegraphics[width=0.3\linewidth]{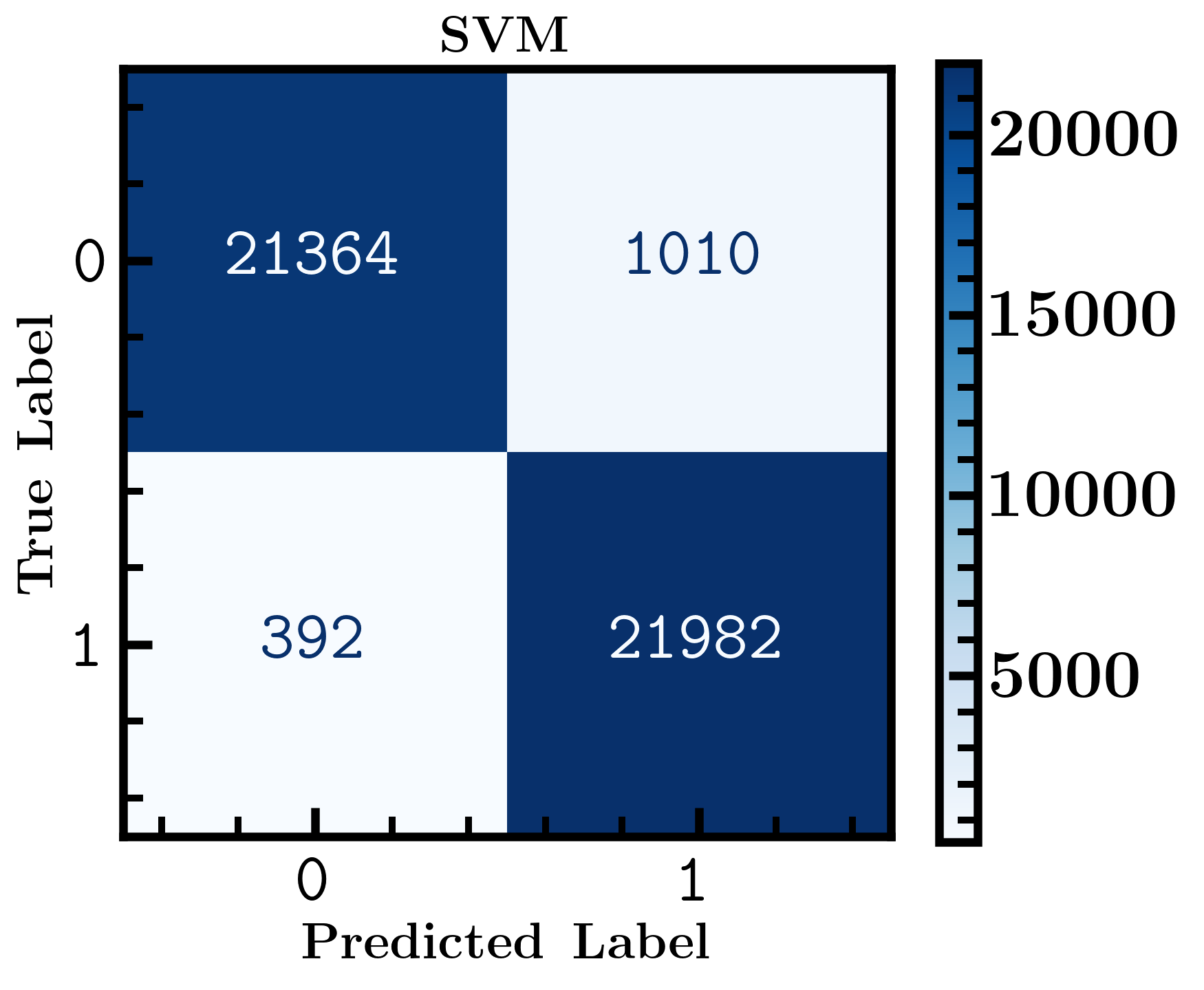}
    \includegraphics[width=0.3\linewidth]{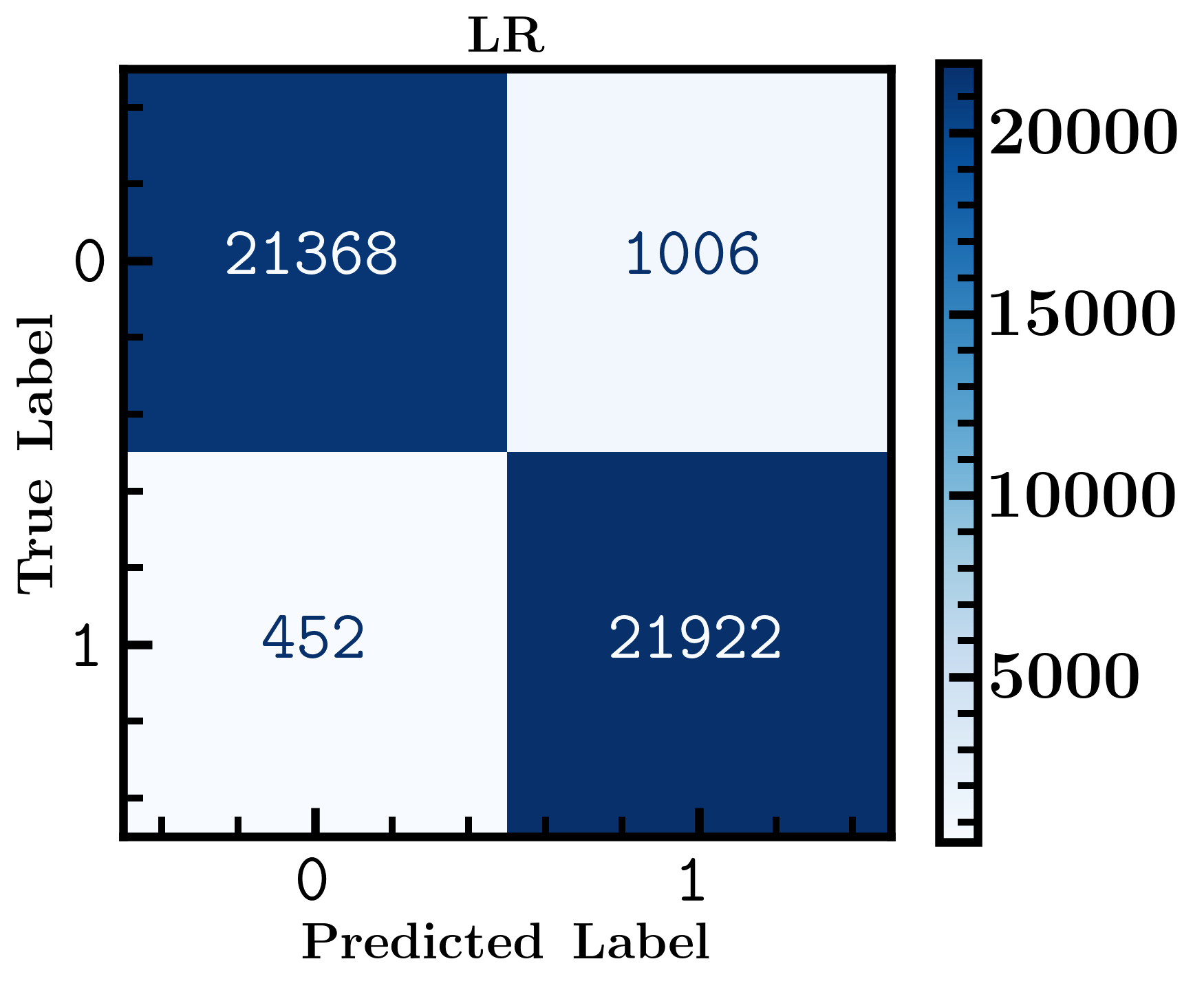}
    \includegraphics[width=0.3\linewidth]{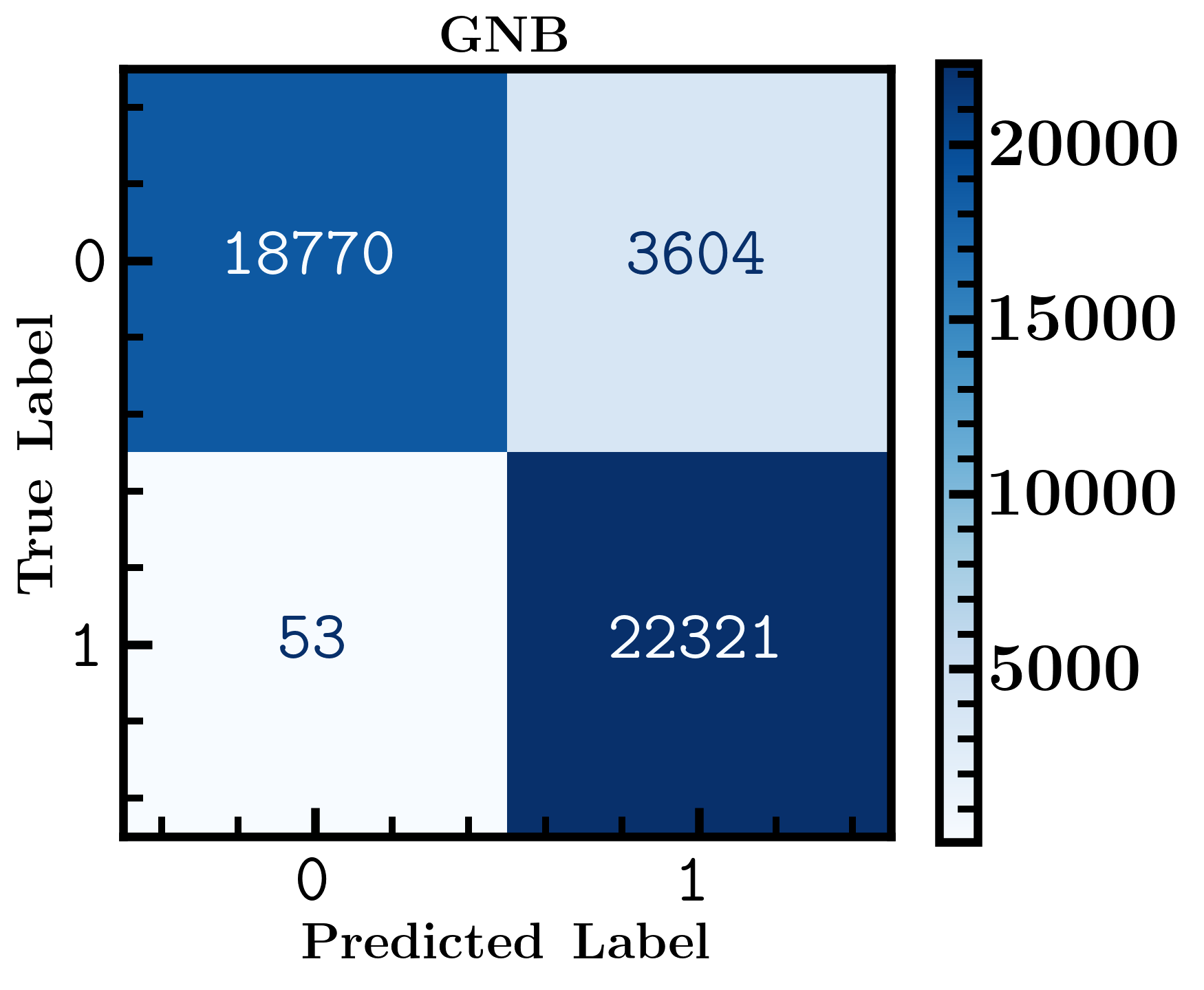}
    \caption{Confusion matrix for different classifiers with $\alpha=+5$}
    \label{fig:conf+5}
\end{figure*}
\begin{figure*}
    \centering
    \includegraphics[width=0.3\linewidth]{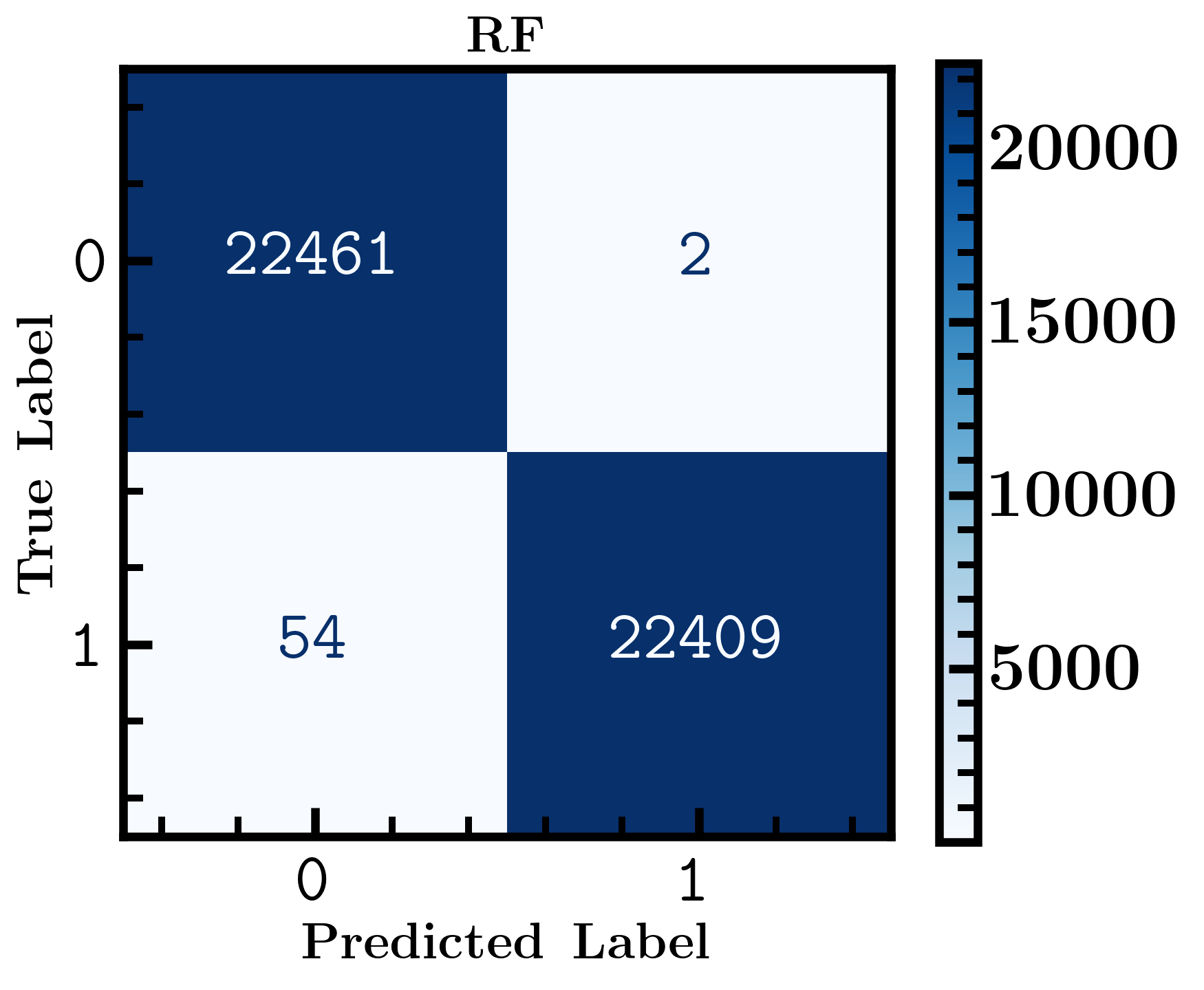}
    \includegraphics[width=0.3\linewidth]{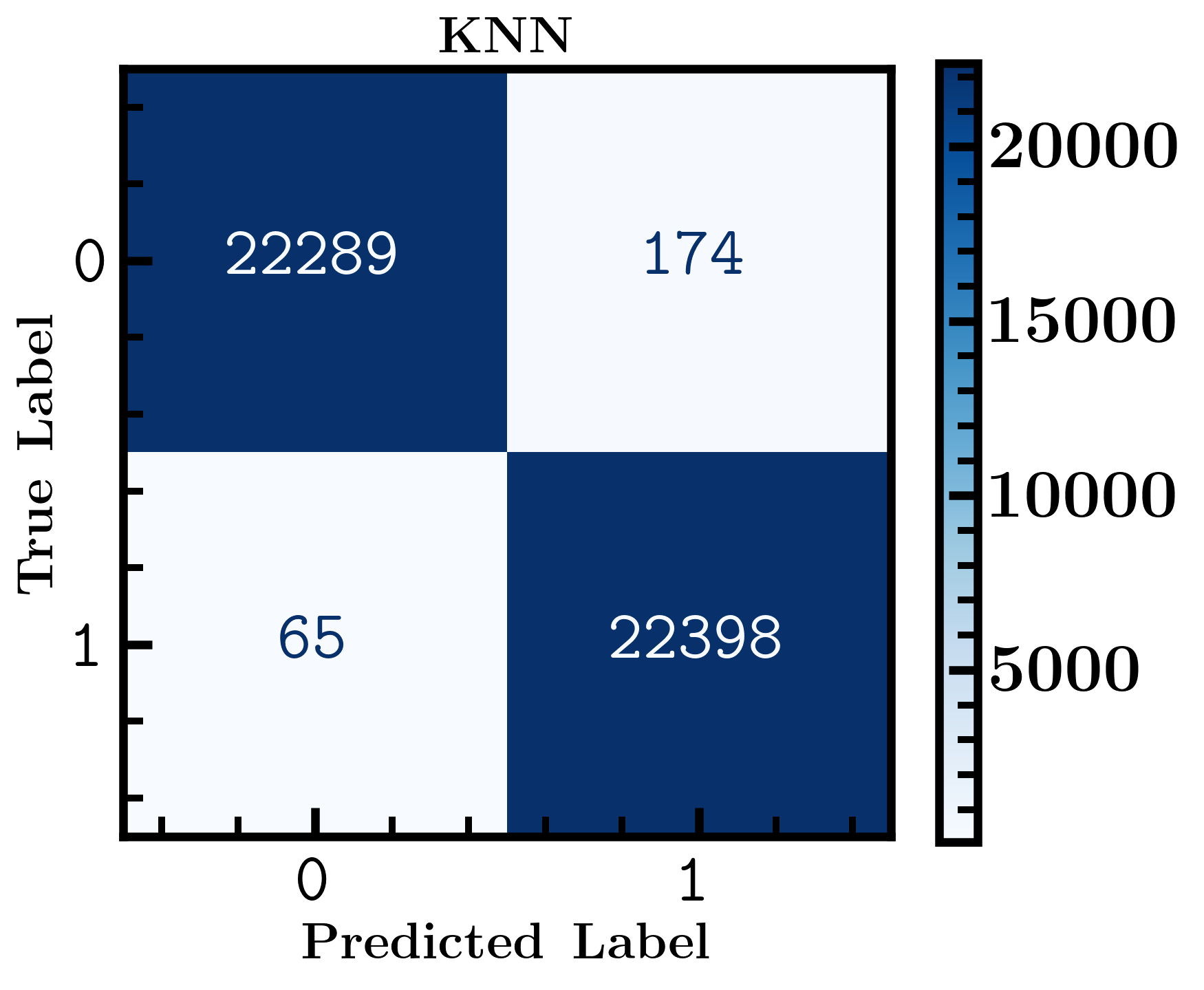}
    \includegraphics[width=0.3\linewidth]{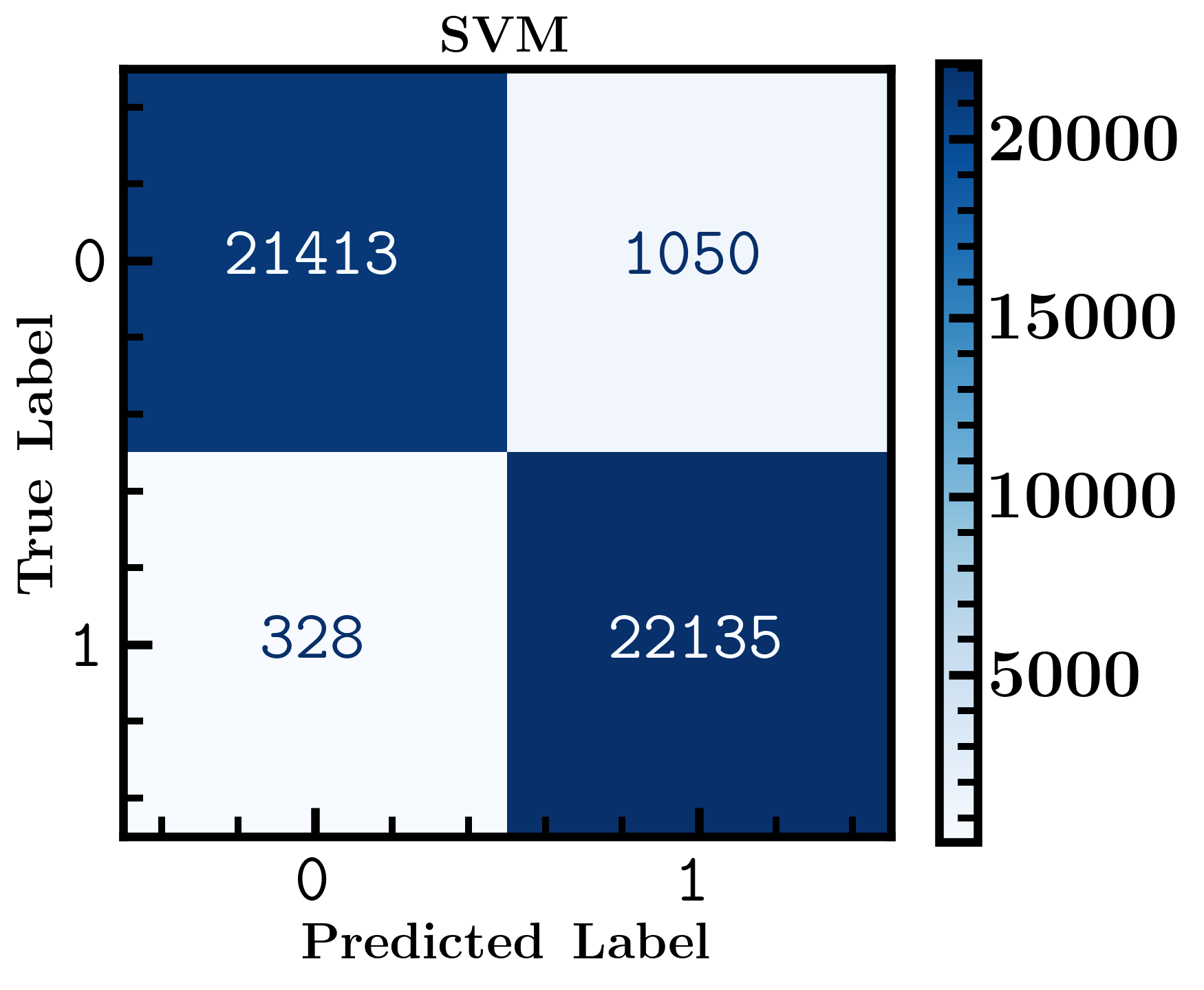}
    \includegraphics[width=0.3\linewidth]{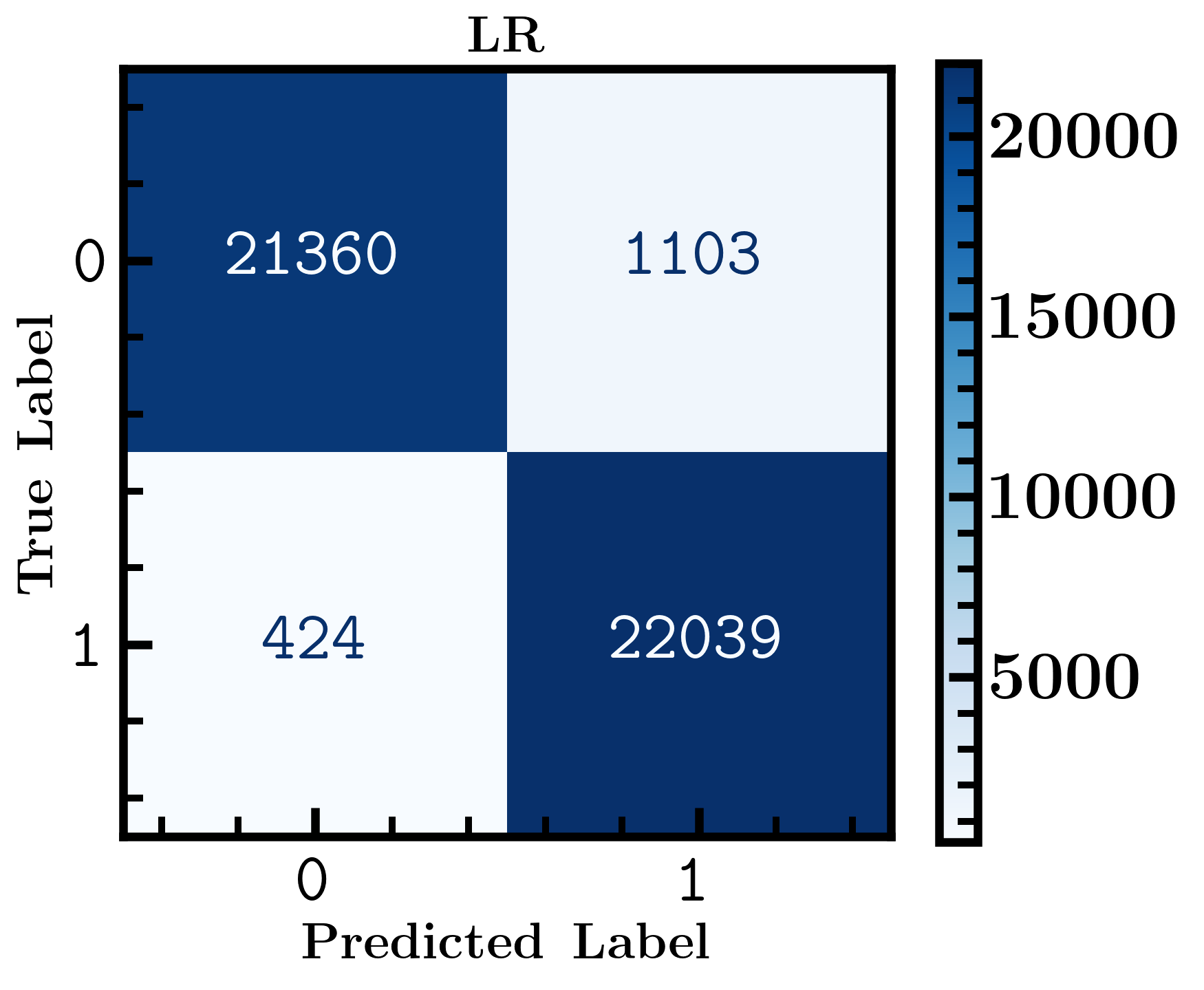}
    \includegraphics[width=0.3\linewidth]{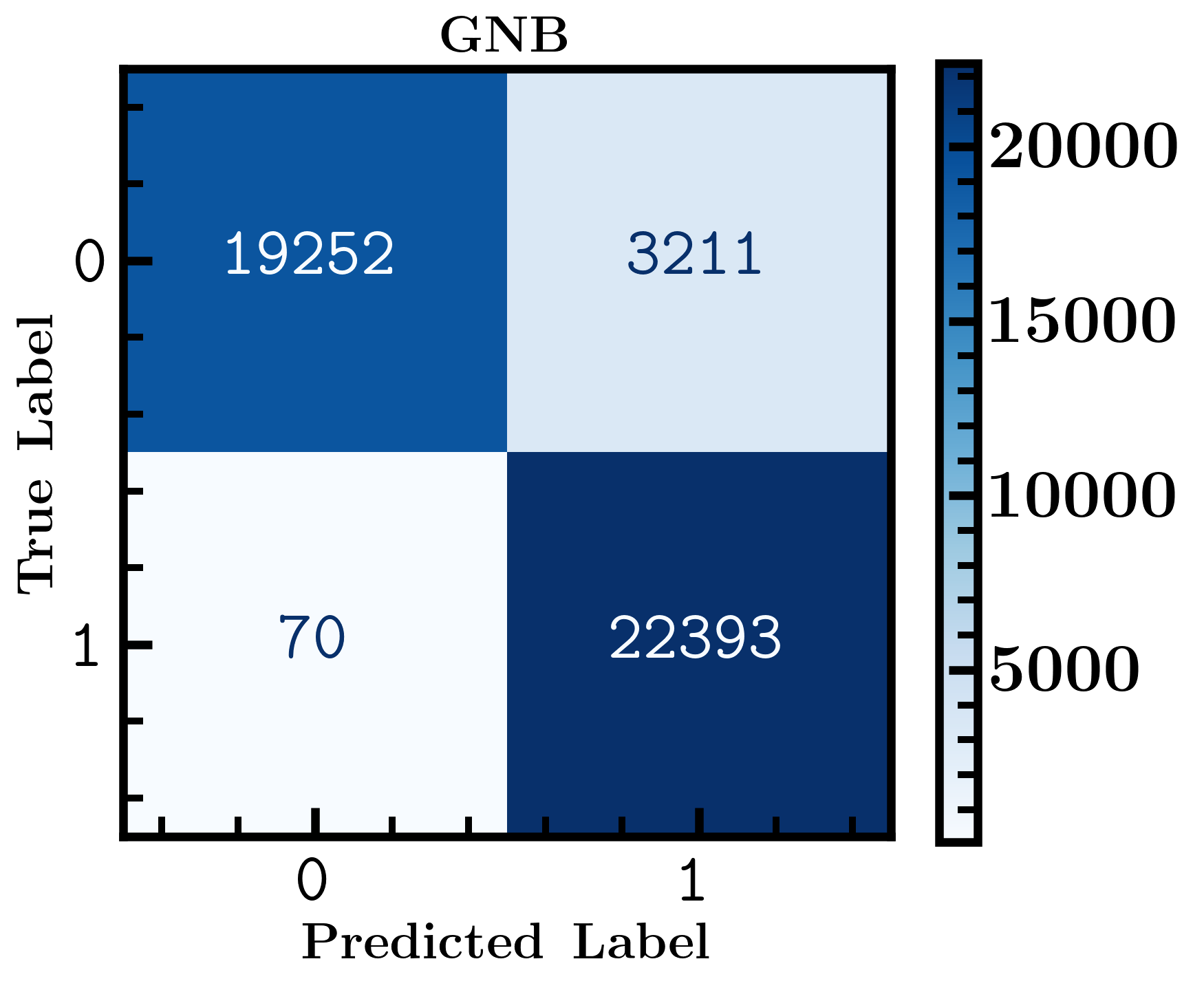}
    \caption{Same as Fig. \ref{fig:conf+5} but for $\alpha=0$}
    \label{fig:conf0}
\end{figure*}

\subsection{Machine learning classification of observationally viable neutron stars}

To investigate whether the macroscopic NS observables contain sufficient information to identify the underlying EMSG coupling, we adopted a two-stage supervised machine-learning framework.
\\
In the first stage, every NS model generated from the approximately $10,000$ EOSs was compared with the available observational constraints in the mass--radius plane. Each configuration was assigned a binary label according to whether its mass--radius pair satisfies the current astrophysical observations. Models consistent with the observational constraints were assigned the label $y=1$, whereas the remaining models were assigned $y=0$. Since the original dataset exhibited class imbalance, the samples were balanced prior to training to prevent the classifiers from favouring the majority class.
\\
Only the NS configurations classified as observationally viable ($y=1$) were retained for the second stage of the analysis. This filtering procedure removes physically excluded models and ensures that the subsequent classification focuses exclusively on neutron stars compatible with current observations.
\\
The filtered dataset was subsequently used to perform a three-label classification corresponding to the three representative EMSG coupling sectors, $\alpha=
\left\{
+5.01\times10^{-38},
0,
-5.01\times10^{-38}
\right\}
\ \mathrm{erg^{-1}\,cm^{3}}$. The input features consisted of the macroscopic observables
$(M,R,\Lambda,f)$,
while the target labels correspond to the three values of the EMSG coupling parameter ($\alpha$). Five supervised machine-learning algorithms, namely Random Forest (RF), K-Nearest Neighbours (KNN), Support Vector Machine (SVM), Logistic Regression (LR), and Gaussian Naive Bayes (GNB), were trained and evaluated using the dataset.
\\
The classification performance is summarized in Table~\ref{tab:APRF} and Fig.~\ref{fig:APRF}. Among all the considered algorithms, the Random Forest classifier achieved the highest predictive performance, yielding an overall accuracy of approximately $99.85\%$, together with precision, recall, and F$_1$-score values exceeding $99.8\%$ for all three EMSG sectors. The KNN classifier also performed remarkably well, achieving accuracies above $99\%$, indicating that observationally allowed NS configurations corresponding to different EMSG sectors occupy well-separated regions in the multidimensional feature space.
\\
Support Vector Machine and Logistic Regression produced accuracies close to $97\%$, demonstrating that the EMSG classes remain distinguishable even with comparatively simple decision boundaries. Although Gaussian Naive Bayes exhibited relatively lower accuracy, it nevertheless correctly classified the majority of NS models. The comparatively lower performance is expected because the NS observables are strongly correlated through the stellar structure equations, partially violating the conditional independence assumption underlying the Gaussian Naive Bayes algorithm.
\\
Overall, these results demonstrate that once observational constraints are imposed, the macroscopic NS observables retain sufficient information to distinguish different EMSG coupling strengths with exceptionally high confidence. The excellent classification performance further indicates that EMSG introduces systematic modifications to NS structure that are readily captured by modern supervised learning techniques.

\subsection{Confusion matrix analysis}

\begin{figure*}[htbp]
    \centering
    \includegraphics[width=0.3\linewidth]{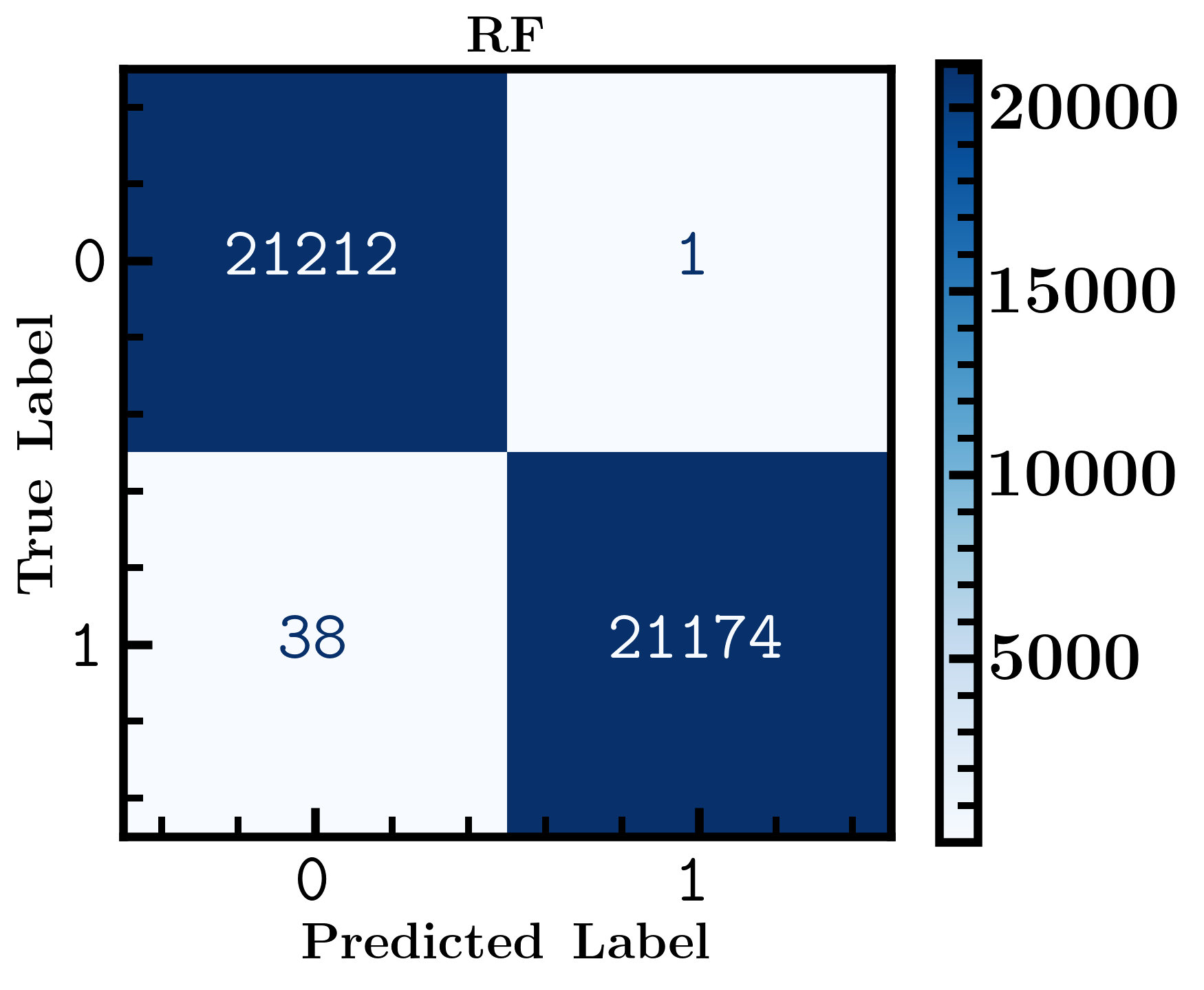}
    \includegraphics[width=0.3\linewidth]{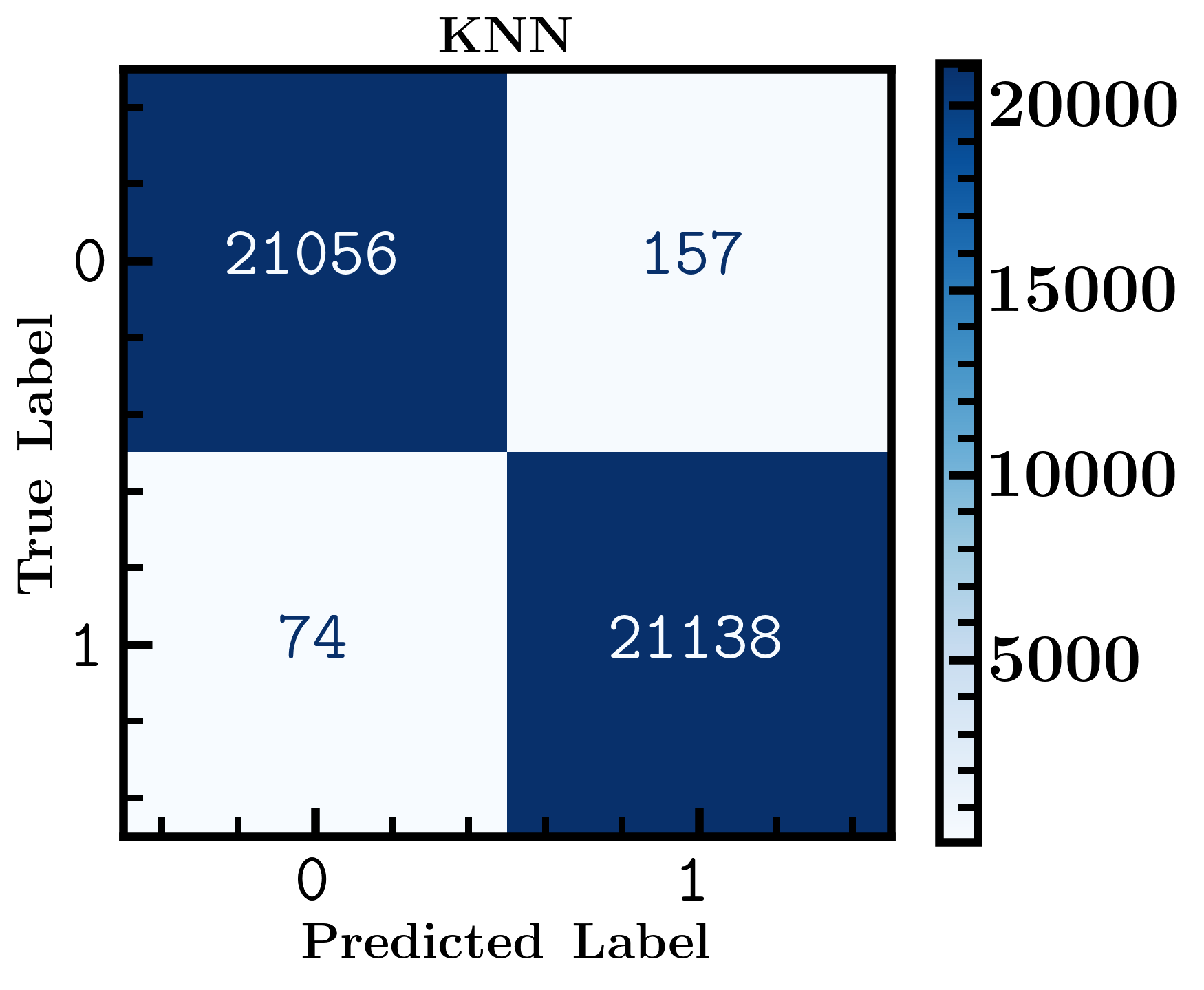}
    \includegraphics[width=0.3\linewidth]{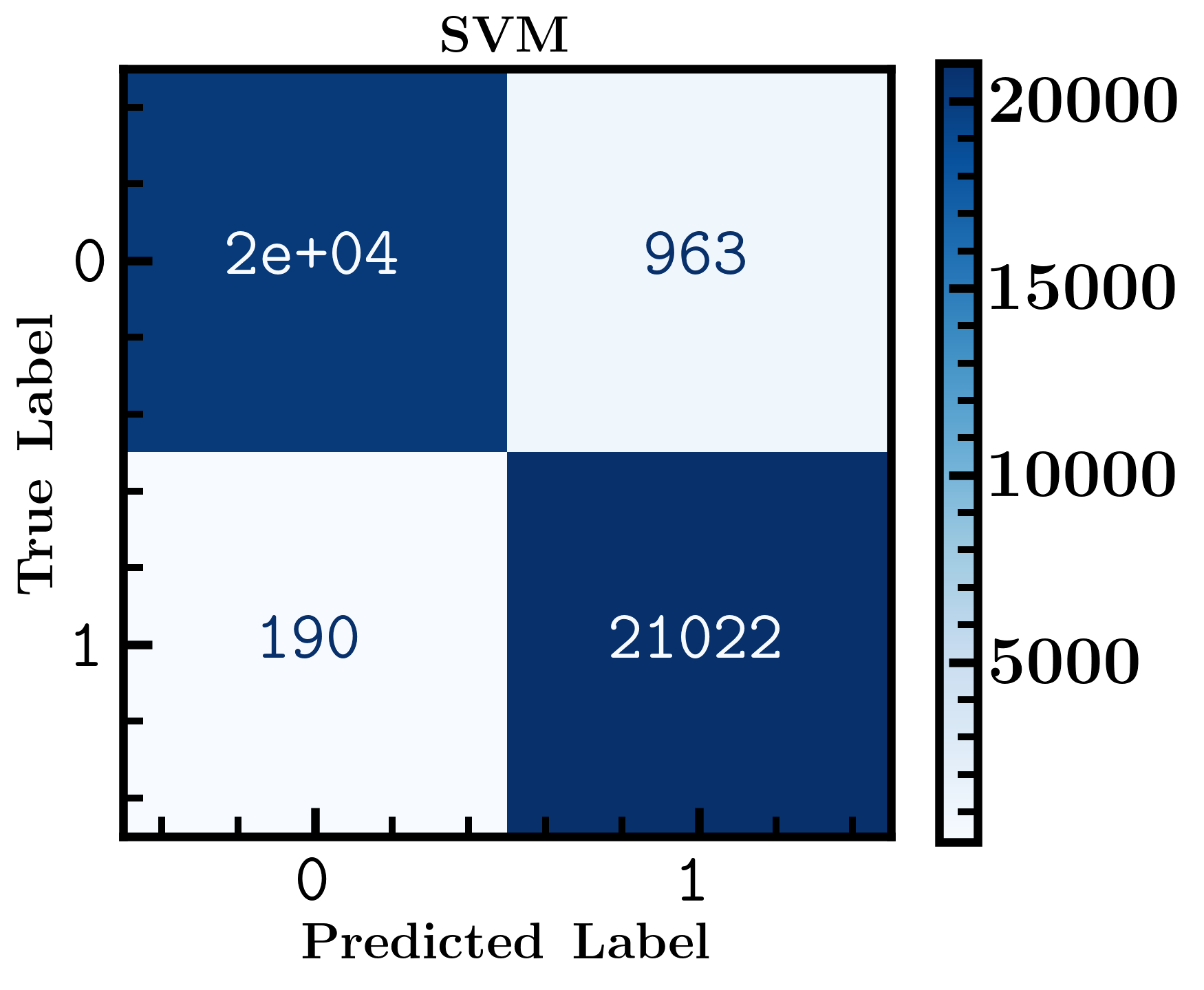}
    \includegraphics[width=0.3\linewidth]{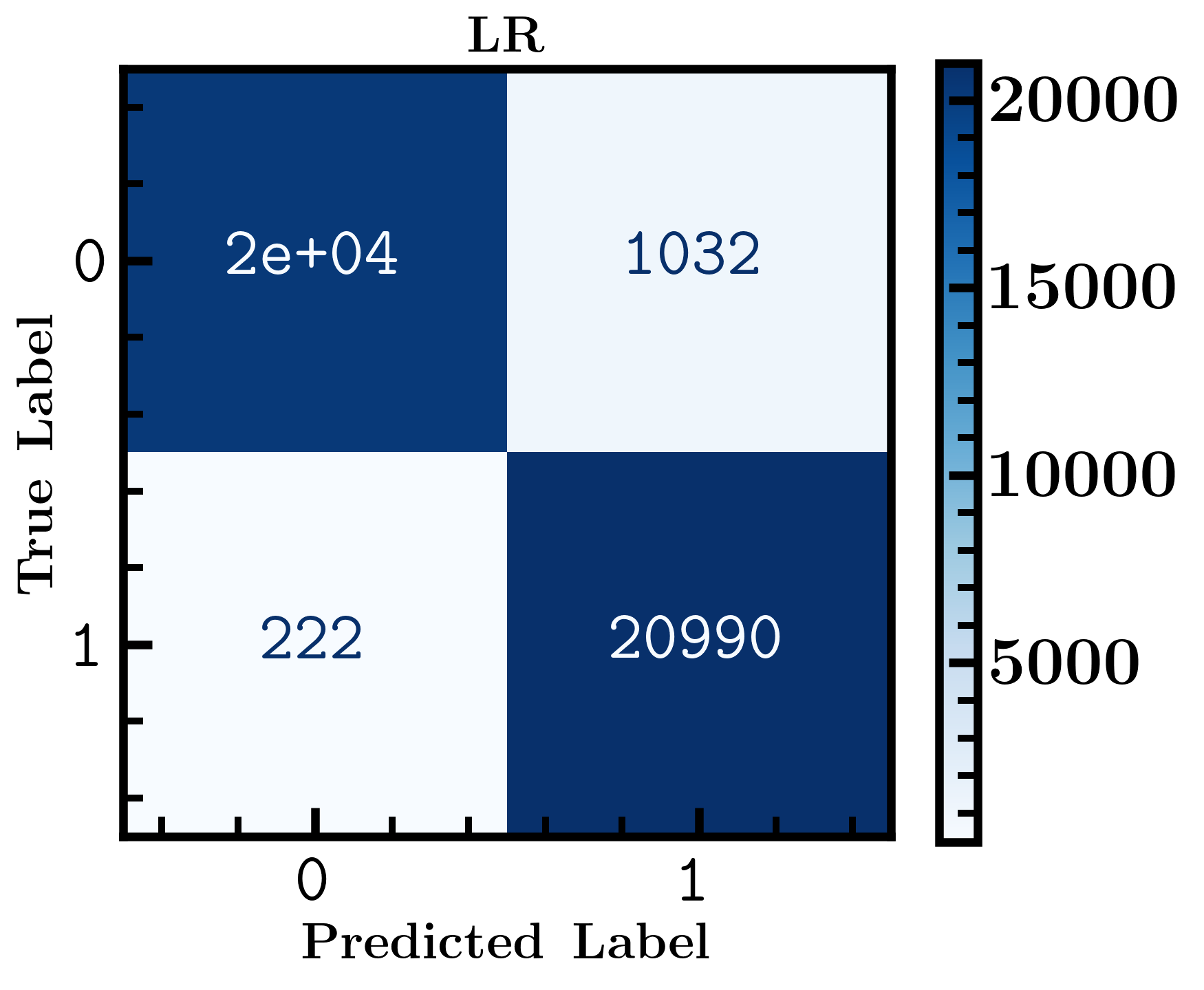}
    \includegraphics[width=0.3\linewidth]{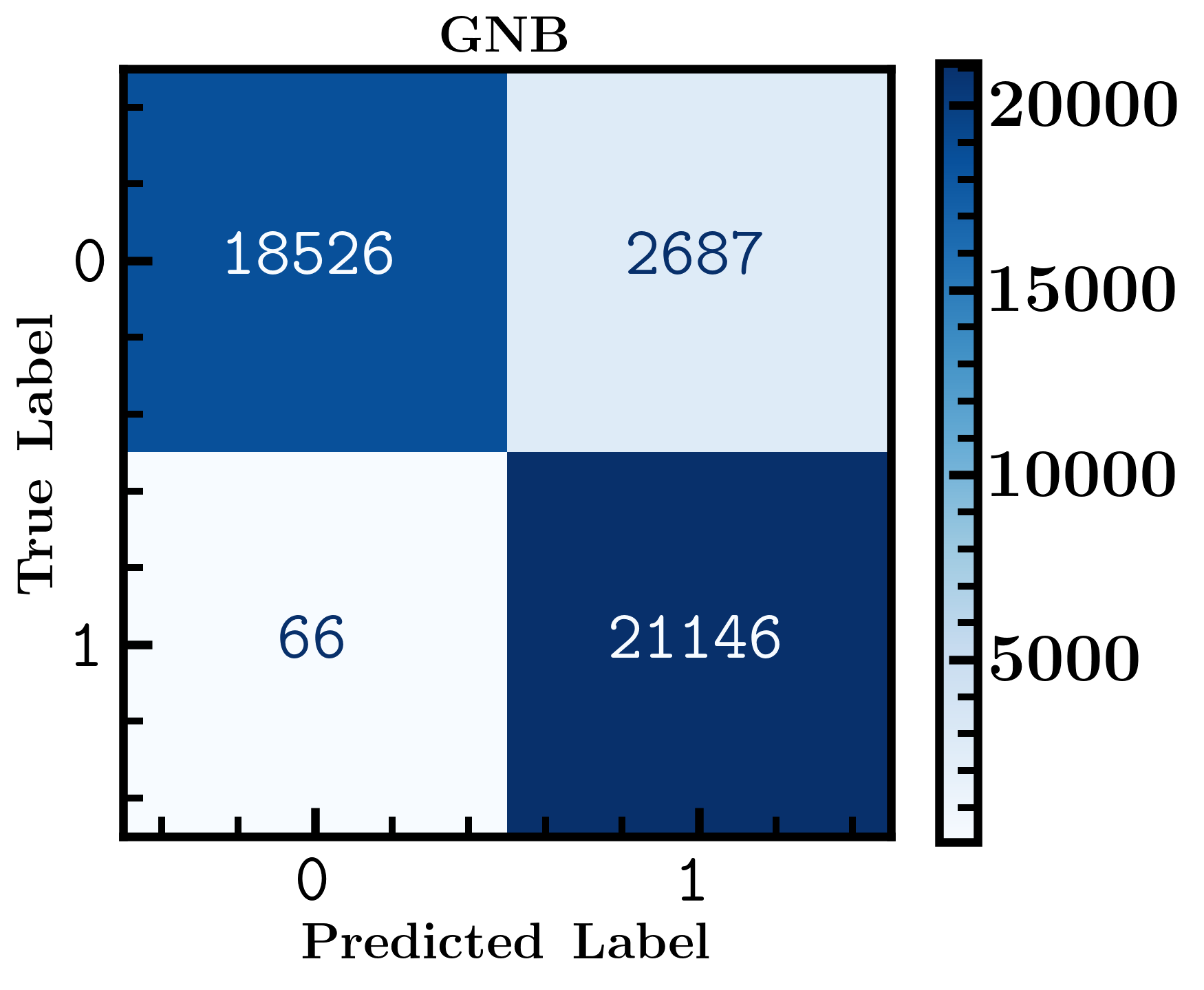}
    \caption{Same as Fig. \ref{fig:conf+5} but for $\alpha=-5$}
    \label{fig:conf-5}
\end{figure*}

To further evaluate the robustness of the machine-learning models, we analysed the confusion matrices corresponding to the binary classification, as shown in Figs.~\ref{fig:conf+5}, \ref{fig:conf0}, \ref{fig:conf-5}. The confusion matrices provide detailed information regarding correctly classified samples and the misclassification rates between different EMSG sectors.
\\
The Random Forest classifier exhibits an almost perfectly diagonal confusion matrix, indicating that nearly all observationally viable NS models are assigned to their correct EMSG class. The numbers of false positives and false negatives remain extremely small, explaining the excellent values of accuracy, precision, recall, and F$_1$-score obtained by the classifier. The KNN model displays a similarly strong diagonal structure, confirming that the EMSG sectors occupy distinct regions in the multidimensional feature space.
\\
SVM and Logistic Regression show slightly larger off-diagonal components, corresponding to a modest increase in confusion between neighbouring EMSG sectors. Gaussian Naive Bayes exhibits the largest off-diagonal entries among the five classifiers, consistent with its comparatively lower performance metrics.
\\
Overall, the confusion-matrix analysis confirms that the combination of NS mass, radius, tidal deformability, moment of inertia, binding energy, and oscillation frequency provides a highly discriminative feature space for identifying the underlying EMSG coupling parameter. The extremely low misclassification rates obtained by the Random Forest classifier demonstrate that machine learning offers a reliable framework for distinguishing modified-gravity signatures using observable NS properties.


\section{Conclusions}
\label{con}
We have investigated whether the macroscopic structure of NS can retain identifiable signatures of Energy-Momentum Squared Gravity (EMSG) after imposing astrophysically motivated constraints. By solving the modified Tolman–Oppenheimer–Volkoff equations over a large set of $10,000$ nuclear equations of state, we find that the EMSG coupling produces systematic changes in NS structure. In particular, positive values of $\alpha$ enhance the effective pressure support and allow larger maximum masses and radii, whereas negative values lead to more compact configurations with reduced maximum masses. The corresponding changes in compactness are also reflected in the tidal deformability and fundamental $f$-mode frequency. Importantly, these effects are not confined to models already excluded by observations: a subset of EMSG configurations remains compatible with the observational constraints considered in this work while retaining measurable departures from the GR predictions. This indicates that observational consistency alone does not necessarily rule out modified-gravity signatures in NS observables.

The ML analysis provides a complementary result. After restricting the dataset to observationally viable configurations, the multidimensional set of observables $(M,R,\Lambda,f)$ remains sufficiently structured to distinguish the three representative EMSG sectors, $\alpha=\{-5.01,0,+5.01\}\times10^{-38}\,\mathrm{erg}^{-1}\mathrm{cm}^{3}$. Among the five classifiers considered, Random Forest gives the strongest performance, with an overall accuracy of approximately $99.85\%$ and precision, recall, and F1-scores above $99.8\%$ for the individual sectors, while K-Nearest Neighbours also exceeds $99\%$ accuracy. The nearly diagonal confusion matrices show that the excellent performance is not simply a consequence of the overall class statistics; configurations associated with the different coupling sectors occupy distinct regions of the observable space.

The main implication is therefore not that ML has replaced the physical modelling of NS, but that it can expose structure in the observable parameter space that is difficult to characterize using a single stellar relation. Once the observationally excluded configurations are removed, the remaining variations in mass, radius, tidal deformability, and oscillation frequency still carry information about the underlying EMSG sector. The result provides evidence that the imprint of the matter–gravity coupling can survive astrophysical filtering and may therefore be relevant for future multi-messenger tests of gravity.

Several limitations should, however, be kept in mind. First, the present analysis is a classification study performed for three discrete and representative values of $\alpha$ rather than a continuous inference of the EMSG coupling. Consequently, the reported classification accuracy should not be interpreted as a direct measurement of $\alpha$ from observational data. Second, the training and testing data are generated from theoretical NS models, and the present analysis does not yet incorporate observational uncertainties and correlated measurement errors in $(M, R,\Lambda,f)$. Third, using a large EOS ensemble reduces sensitivity to any single nuclear model, but the EOS space considered here does not exhaust the full theoretical uncertainty in the high-density equation of state. Finally, the fundamental $f$-mode frequency is calculated within the Cowling approximation, in which metric perturbations are neglected; this introduces an additional modelling limitation when connecting the predicted oscillation frequencies to future gravitational-wave measurements.

These limitations motivate several natural extensions. A particularly important next step is to replace the discrete classification of $\alpha$ by a regression or Bayesian inference framework capable of recovering a posterior distribution for the coupling parameter while propagating uncertainties in the EOS and in the measured NS observables. The present observable set can also be augmented, where appropriate, by additional quantities such as the moment of inertia and binding-energy-related observables, allowing us to test whether they provide independent information on the EMSG coupling rather than simply correlated information. A further extension will be to include realistic observational uncertainties and synthetic measurement errors in the training and testing samples, thereby testing whether the high separability found in the idealized theoretical dataset survives under realistic data quality. The inclusion of more general EMSG couplings, a continuous range of $\alpha$, and alternative modified-gravity theories would then allow the framework to address model discrimination rather than classification within a single theory.

Overall, the results establish a proof of principle that NS observables can retain distinguishable signatures of EMSG even within an observationally viable region of the parameter space. The combination of relativistic stellar modelling, nuclear-EOS uncertainty, and data-driven inference therefore offers a useful route toward testing gravity with neutron stars. As measurements of masses, radii, tidal properties, and NS oscillations become increasingly precise, such a framework can be developed from a classification tool into a statistically controlled method for constraining deviations from General Relativity.

\section*{Acknowledgement}
S.G. and D.D. acknowledge the National Institute of Technology Rourkela for fellowship support.  P.M would like to thank BITS Pilani, K K Birla Goa campus for the fellowship support; Subhadip Sau and Ayush Hazarika for the fruitful discussions on Machine Learning analysis.


\bibliography{reference}
\end{document}